\documentclass{aa}

\usepackage{graphicx}
\usepackage{subcaption}
\usepackage{txfonts}
\usepackage{siunitx}
\usepackage{booktabs}
\usepackage[flushleft]{threeparttable}
\usepackage{comment}
\usepackage{makecell}
\usepackage[normalem]{ulem}
\usepackage{subcaption}

\PassOptionsToPackage{draft}{hyperref}
\usepackage{xspace}

\usepackage[xindy, toc, hyperfirst=false, nolist, nostyles, sanitize={name=false,description=false,symbol=false}]{glossaries}
\glsdisablehyper
\usepackage[hyperref,x11names, table]{xcolor}

\newcommand{\msun}{\hbox{$M_{\odot}$}}

\newglossaryentry{vrad}{name={radial velocity~}, text={radial velocity}, symbol={\ensuremath{v_\textrm{rad}}}, description={radial velocity}, sort=vrad}
\newglossaryentry{vrot}{name={stellar rotation~}, name={stellar rotation}, symbol={\ensuremath{v_\textrm{rot}}}, description={radial velocity}, sort=vrot}

\newcommand{\kms}{\ensuremath{\textrm{km}~\textrm{s}^{-1}}}

\newcommand{\xray}{X-ray}

\newglossaryentry{angstrom}{name=\AA, description={unit of length $10^{-10}$\,m}, sort=angstrom}
\newglossaryentry{nir}{name=NIR,description={near infrared},first = {near infrared (NIR)}}
\newglossaryentry{psf}{name=PSF,description={point-spread function},first = {point-spread function (PSF)}}
\newglossaryentry{fwhm}{name=FWHM,description={Full Width Half Maximum},first = {FWHM}}
\newglossaryentry{rms}{name=RMS,description={Root Mean Square},first = {RMS}}
\newglossaryentry{signalnoise}{name=S/N,description={signal to noise}}
\newglossaryentry{uv}{name=UV,description={ultra violet},first = {ultra violet (UV)}}
\newglossaryentry{halpha}{name=\ensuremath{\textrm{H}\alpha}, description={First line of the Balmer series at 6563\,\AA}, sort=halpha}
\newglossaryentry{mgb}{name={Mg \textsc{i} b}, description={Triplet at 5167\,\AA, 5173\,\AA and 5184\,\AA}}
\newglossaryentry{sobolevapprox}{name={Sobolev approximation}, description={Lines are approximation with an infinitley thin interaction region \citep[e.g. no broadening][]{1960mes..book.....S}}, first={Sobolev approximation }}
\newglossaryentry{radeq}{name={radiative equilibrium}, description={The net flux of energy between matter and radiation field is zero}}
\newglossaryentry{nebularapprox}{name={nebular approximation}, description={Assumes that the plasma condition are controlled by a central radiation source. The radiation field decreases with the distance to the source by geometrical dilution. See \citet{1978stat.book.....M} for details}}
\newglossaryentry{modnebularapprox}{name={modified nebular approximation}, description={In contrast to \gls{nebularapprox} where only geometrical dilution is taken into account, the modified nebular approximation also takes dilution by other radiative processes into account }, first={modified nebular approximation}, parent=nebularapprox}
\newglossaryentry{thompsonscat}{name={Thomson scattering}, description={Scattering of photons on low energy electrons}}
\newglossaryentry{lte}{name={LTE}, description={Local Thermodynamic Equilibrium}, first={local thermodynamic equilibrium (LTE)}}
\newglossaryentry{lsr}{name={LSR}, description={Local Standard of Rest}, first={\textit{local standard of rest} (LSR)}}
\newglossaryentry{mc}{name={MC}, description={Monte Carlo}, first={\textit{Monte Carlo} (MC)}}
\newglossaryentry{wcs}{name={WCS}, description={world coordinate system}, first={world coordinate system (WCS)}}
\newglossaryentry{cmf}{name=CMF, text=CMF, first=Comoving Frame (CMF henceforth), description={Comoving Frame}}

\newglossaryentry{uvoir}{name=UVOIR, text=UVOIR, first=UV/optical/Near-IR (UVOIR), description={UV/optical/Near-IR}}

\newglossaryentry{sfit}{name=SFIT, text=\textsc{sfit}, description={spectral fitting program for hot stars \citep{2001A&A...376..497J}}, first={\textsc{sfit} \citep{2001A&A...376..497J}}}
\newglossaryentry{iraf}{name=IRAF, text=\textsc{iraf}, description={Image Reduction and Analysis Facility maintained by NOAO}, first={\textsc{iraf}\protect\footnote{IRAF: the Image Reduction and Analysis Facility is distributed by the National Optical Astronomy Observatory, which is operated by the Association of Universities for Research in Astronomy (AURA) under cooperative agreement with the National Science Foundation (NSF).}}}
\newglossaryentry{pyraf}{name=PyRAF, text=\textsc{PyRAF}, description={Python wrap of \gls{iraf} maintained by STSCI}, first=\textsc{PyRAF} \protect\footnote{PyRAF is a product of the Space Telescope Science Institute, which is operated by AURA for NASA.}}
\newglossaryentry{astropy}{name=ASTROPY, text=\textsc{astropy}, description=\textsc{astropy} framework, first = \textsc{astropy} \citep{2013A&A...558A..33A}}
\newglossaryentry{numpy}{name=NUMPY, text=\textsc{numpy}, description=\textsc{numpy} framework, first = \textsc{numpy} \citep{walt2011numpy}}
\newglossaryentry{scipy}{name=SCIPY, text=\textsc{scipy}, description=\textsc{scipy} framework, first = \textsc{scipy} \citep{Jones:2001fk}}
\newglossaryentry{matplotlib}{name=matplotlib, text=\textsc{matplotlib}, description=\textsc{matplotlib} framework, first = \textsc{matplotlib} \citep{hunter2007matplotlib}}
\newglossaryentry{pandas}{name=pandas, text=\textsc{pandas}, description=\textsc{pandas} framework, first = \textsc{pandas} \citep{mckinney2010data}}
\newglossaryentry{ipython}{name=ipython, text=\textsc{ipython}, description=\textsc{ipython} framework, first = \textsc{ipython} \citep{perez2007ipython}}
\newglossaryentry{jupyter}{name=jupyter, text=\textsc{jupyter}, description=\textsc{jupyter} framework, first = \textsc{jupyter} \citep{kluyver2016jupyter,perez2015project,ragan2014jupyter}}
\newglossaryentry{aplpy}{name=aplpy, text=\textsc{aplpy}, description=\textsc{aplpy} framework, first = \textsc{aplpy} \citep{2012ascl.soft08017R}}
\newglossaryentry{nltk}{name=nltk, text=\textsc{nltk}, description=\textsc{nltk} framework, first = Natural Language ToolKit \citep[\textsc{NLTK};][]{bird2009natural}}
\newglossaryentry{scikit-learn}{name=scikit-learn, text=\textsc{scikit-learn}, description=\textsc{scikit-learn} framework, first = \textsc{scikit-learn} \citep[][]{scikit-learn}}
\newglossaryentry{scikit-image}{name=scikit-image, text=\textsc{scikit-image}, description=\textsc{scikit-image} framework, first = \textsc{scikit-image} \citep[][]{scikit-image}}
\newglossaryentry{moog}{name=MOOG,text={\textsc{moog}}, description={spectral synthesis software \citep{1973ApJ...184..839S}}, first={\textsc{Moog} \citep{1973ApJ...184..839S}}}
\newglossaryentry{atlas9}{name=ATLAS9,description={grid of stellar atmospheres \citep{2004astro.ph..5087C}}, first={ATLAS9 \citep{2004astro.ph..5087C}}}
\newglossaryentry{vald}{name=VALD,description={Vienna Atomic Line Database \citep{2000BaltA...9..590K}}, first={Vienna Atomic Line Database \citep[VALD;][]{2000BaltA...9..590K}}}
\newglossaryentry{sextractor}{name=SExtractor, text=\textsc{SExtractor}, description={Source Extractor photometry program \citep{1996A&AS..117..393B}}, first={\textsc{SExtractor} \citep{1996A&AS..117..393B}}}

\newglossaryentry{swarp}{name=SWarp, text=\textsc{SWarp}, description={SWarp \citep{2002ASPC..281..228B}}, first={\textsc{SWarp} \citep{2002ASPC..281..228B}}}
\newglossaryentry{astrometry.net}{name=astrometry.net, text=\textsc{astrometry.net}, description={\textsc{astrometry.net} \citep{2010AJ....139.1782L}} first={\textsc{astrometry.net} \citep{2010AJ....139.1782L}}}

\newglossaryentry{astrodrizzle}{name=AstroDrizzle, text=\textsc{AstroDrizzle}, description={AstroDrizzle \citep{2012drzp.book.....G}}, first={\textsc{AstroDrizzle} \citep{2012drzp.book.....G}}}

\newglossaryentry{idl}{name=IDL,text={\textsc{idl}}, description={Interactive Data Language}}
\newglossaryentry{makee}{name=MAKEE,text=\textsc{makee}, description={MAuna Kea Echelle Extraction by Tom Barlow available}}% at \verb+http://spider.ipac.caltech.edu/staff/tab/makee/index.html+}}
\newglossaryentry{minuit}{name=MINUIT,text={\textsc{minuit}}, description={collection of numerical optimization tools \citep{James:1975dr}}}
\newglossaryentry{migrad}{name=MIGRAD,text={\textsc{migrad}}, description={numerical gradient optimization tools - part of \gls{minuit}}}
\newglossaryentry{dolphot}{name=DOLPHOT, text=\textsc{dolphot}, description=photometry package for HST, first=\textsc{dolphot} \citep{2000PASP..112.1383D}}
\newglossaryentry{synphot}{name=synphot, text={\textsc{synphot}}, description={synthetic photometry package from STSCI}, first={\textsc{synphot}\protect\footnote{\textsc{synphot} is a product of the Space Telescope Science Institute, which is operated by AURA for NASA.}}}
\newglossaryentry{chianti}{name=CHIANTI, text=CHIANTI, description= CHIANTI Database 7.1, first =CHIANTI 7.1 \citep{1997A&AS..125..149D,2012ApJ...744...99L}}
\newglossaryentry{synpp}{name=SYNPP, text=SYN++, description= SYN++ software, first =SYN++ \citep{2011PASP..123..237T}}
\newglossaryentry{tardis}{name=TARDIS, text=\textsc{tardis}, description= TARDIS MC code, first = {\textsc{tardis} \citep{2014MNRAS.440..387K}}}

\newglossaryentry{artis}{name=ARTIS, text=\textsc{artis}, description= ARTIS MC code, first = \textsc{artis} \citep{2009MNRAS.398.1809K}}
\newglossaryentry{cmfgen}{name=CMFGEN, text=\textsc{cmfgen}, description=CMFGGEn radiative transfer code, first = \textsc{cmfgen} \citep{1998ApJ...496..407H}}
\newglossaryentry{sedona}{name=SEDONA, text=\textsc{sedona}, description= Sedona MC code, first = \textsc{sedona} \citep{2006ApJ...651..366K}}
\newglossaryentry{mlmc}{name=MLMC, text=ML93, description= Mazzali Lucy Monte Carlo, first ={Mazzali \& Lucy (1993, ML93) code}}
\newglossaryentry{starkit}{name=STARKIT, text=\textsc{starkit}, description= TARDIS MC code, first = {\textsc{starkit} \citep{wolfgang_kerzendorf_2015_28016}}}

\newglossaryentry{pyne}{name=PYNE, text=\textsc{pyne}, description= PYNE code, first = {\textsc{pyne} \citep{Scopatz2012a}}}
\newglossaryentry{multinest}{name=MULTINEST, text=\textsc{MultiNest}, description=MultiNest, first={\textsc{MultiNest} \citep{2009MNRAS.398.1601F}}}
\newglossaryentry{wsynphot}{name=WSYNPHOT, text=\textsc{wsynphot}, description=Wsynphot, first={\textsc{wsynphot}\protect\footnote{\protect\url{https://github.com/wkerzendorf/wsynphot}}}}
\newglossaryentry{specutils}{name=SPECUTILS, text=\textsc{specutils}, description=specutils, first={\textsc{specutils} \protect\footnote{\protect\url{https://github.com/astropy/specutils}}}}
\newglossaryentry{ads}{name=ADS ,description=ADS, first={NASA Astrophysics Data System (ADS) \citep{2000A&AS..143...41K}}}

\newglossaryentry{2mass}{name=2MASS,description={Two Micron All Sky Survey \citep{2006AJ....131.1163S}}, first={Two Micron All Sky Survey \citep{2006AJ....131.1163S}}}
\newglossaryentry{nomad}{name=NOMAD,first={Naval Observatory Merged Astrometric Dataset \citep[NOMAD; ][]{2005yCat.1297....0Z}}, description={Naval Observatory Merged Astrometric Dataset}}
\newglossaryentry{wifes}{name=WIFES, text=\textsc{WiFeS}, first={\textsc{WiFeS} \citep{2007Ap&SS.310..255D}},  description={Wide Field Spectrograph - \gls{ifu} mounted on the 2.3\,m telescope at Siding Spring Observatory}}
\newglossaryentry{scp}{name=SCP,description={Supernova Cosmology Project, led by Saul Perlmutter}, first={Supernova Cosmology Project (SCP)}}
\newglossaryentry{hzsns}{name=HZSNS,description={High Z Supernova Search, led by Brian Schmidt}, first={High Z Supernova Search (HZSNS)}}
\newglossaryentry{vlt}{name=VLT,description={Very Large Telescope located on Cerro Paranal (Chile)}, first={Very Large Telescope (VLT)}}
\newglossaryentry{flames}{name=FLAMES,description={Multi-object, intermediate and high resolution spectrograph mounted on the  \gls{vlt}}}
\newglossaryentry{hires}{name=HIRES, description={High Resolution Echelle Spectrometer mounted on the Keck Telescope}, first={High Resolution Echelle Spectrometer \citep[HIRES;][]{1994SPIE.2198..362V}}}
\newglossaryentry{lris}{name=LRIS,description={Low Resolution Imaging Spectrometer mounted on the Keck Telescope}, first={Low-Resolution Imaging Spectrometer \citep[LRIS;][]{Oke95}}}

\newglossaryentry{decam}{name=DECam, description={DECam is a high-performance, wide-field CCD imager mounted at the prime focus of the Blanco 4-m telescope at \gls{ctio}.}, first={Dark Energy Camera \citep[DECam; ][]{2012PhPro..37.1332D,2015AJ....150..150F}}}

\newglossaryentry{essence}{name=ESSENCE,description={The `Equation of State: SupErNovae trace Cosmic Expansion' project \citep[ESSENCE;][]{2002AAS...201.7809G}}, first={`The Equation of State: SupErNovae trace Cosmic Expansion' \citep[ESSENCE;][]{2002AAS...201.7809G}}}
\newglossaryentry{ifu}{name=IFU,description={Optical instrument combining spectrographic and imaging capabilities, used to obtain spatially resolved spectra}, first={Integral Field Unit (IFU)}, firstplural={Integral Field Units (IFUs)}}

\newglossaryentry{besancon}{name=Besan\c{c}on Model, description={Model of stellar population synthesis of the Galaxy, including kinematics.}}%  \verb+http://model.obs-besancon.fr+} }, nonumberlist=true}

\newglossaryentry{int}{name=INT,description={Isaac Newton 2.5\,m Telescope}, first={Isaac Newton 2.5\,m Telescope (INT)}}
\newglossaryentry{iau}{name=IAU,description={International Astronomical Union}, first={IAU}}
\newglossaryentry{chandra}{name=Chandra,description={Chandra \xray\ Observatory (space-based)}}
\newglossaryentry{hst}{name=HST,description={Hubble Space Telescope}}
\newglossaryentry{hst.wfpc2}{name=WFPC2,description={Wide-Field Planetary Camera 2 mounted on the \gls{hst}}, first={Wide-Field Planetary Camera 2 (WFPC2)}}
\newglossaryentry{hst.acs}{name=ACS,description={Advanced Camera for Surveys mounted on the \gls{hst}}, first={Advanced Camera for Surveys (ACS)}}
\newglossaryentry{hst.wfc3}{name=WFC3,description={Wide-Field Camera 3 mounted on the \gls{hst}}, first={Wide-Field Camera 3 (WFC3)}}
\newglossaryentry{hst.cte}{name=CTE, description={charge transfer efficiency (CTE)}, first={charge transfer efficiency \citep[CTE; see ][for a description]{2009acs..rept....1C}}}

\newglossaryentry{snls}{name=SNLS,description={Supernova Legacy Survey \citep{2003AAS...203.8209P}}, first={Supernova Legacy Survey \citep[SNLS;][]{2003AAS...203.8209P}}}
\newglossaryentry{dass}{name=DASS, description={Digitized Astronomy Supernova Survey \citep{1975PASP...87..565C}}, first={Digitized Astronomy Supernova Survey \citep[DASS;][]{1975PASP...87..565C}}}
\newglossaryentry{bait}{name=BAIT, description={Berkley Automatic Imaging Telescope \citep{1993PASP..105.1164R}}, first={Berkley Automatic Imaging Telescope \citep[BAIT;][]{1993PASP..105.1164R}}}
\newglossaryentry{kait}{name=KAIT, description={Katzman Automatic Imaging Telescope \citep{2001ASPC..246..121F}}, first={Katzman Automatic Imaging Telescope \citep[KAIT;][]{2001ASPC..246..121F}}}
\newglossaryentry{loss}{name=LOSS, description={Lick Observatory Supernova Search  \citep{2000AIPC..522..103L}}, first={Lick Observatory Supernova Search \citep[LOSS;][]{2000AIPC..522..103L}}}
\newglossaryentry{ctss}{name=CTSS,description={Cal\'{a}n/Tololo Supernova Survey \citep{1993AJ....106.2392H}}, first={Cal\'{a}n/Tololo supernova survey \citep[CTSS;][]{1993AJ....106.2392H}}}
\newglossaryentry{ctio}{name= CTIO, description={Cerro Tololo Inter-American Observatory}, first={Cerro Tololo Inter-American Observatory (CTIO)}}
\newglossaryentry{ptf}{name=PTF, description={Palomar Transient Factory \citep{2009PASP..121.1334R}}, first={Palomar Transient Factory \citep[PTF;][]{2009PASP..121.1334R}}}
\newglossaryentry{batse}{name=BATSE, description={Burst and Transient Source Experiment mounted on the Compton Gamma Ray Observatory}, first={Burst and Transient Source Experiment (BATSE)}}
\newglossaryentry{bepposax}{name=BeppoSAX, description={\xray\ satellite named in honor of Giuseppe "Beppo" Occhialini}}
\newglossaryentry{rosat}{name=ROSAT, description={short for R\"{o}ntgensatellit}, first={ROSAT}}
\newglossaryentry{hete2}{name=HETE2, description={High Energy Transient Explorer}, first={High Energy Transient Explorer (HETE)}}
\newglossaryentry{ska}{name=SKA, description={Square Kilometre Array}, first={Square Kilometre Array (SKA)}}

\newglossaryentry{gnirs}{name=GNIRS, description={Gemini Near InfraRed Spectrograph mounted on the Gemini North Telescope}}
\newglossaryentry{gmosn}{name=GMOS, description={Gemini Multi Object Spectrograph mounted on the
 Gemini North Telescope}, first={GMOS \citep[Gemini Multi Object Spectrograph;][]{2004PASP..116..425H}}}
\newglossaryentry{swift}{name=Swift, description={Swift Gamma-Ray Burst Mission}}
\newglossaryentry{vla}{name=VLA, description={Very Large Array radio telescope located in North America}, first={Very Large Array (VLA)}}
\newglossaryentry{evla}{name=EVLA, description={Extended Very Large Array radio telescope located in North America}, first={Extended Very Large Array (EVLA)}}
\newglossaryentry{sdss}{name=SDSS, description={Sloan Digital Sky Survey}}
\newglossaryentry{dss}{name=DSS, description={Digitized Sky Survey}}
\newglossaryentry{skymapper}{name=SkyMapper, description={SkyMapper telescope \citep{2007PASA...24....1K}}, first={SkyMapper \citep{2007PASA...24....1K}}}
\newglossaryentry{panstarrs}{name=PanSTARRS, description={Panoramic Survey Telescope \& Rapid Response System \citep{2004SPIE.5489...11K}}, first={Panoramic Survey Telescope \& Rapid Response System \citep[PanSTARRS;][]{2004SPIE.5489...11K}}}
\newglossaryentry{ps1dr1}{name=PS1~DR1, description={Panoramic Survey Telescope \& Rapid Response System \citep{2004SPIE.5489...11K} }, first={Panoramic Survey Telescope \& Rapid Response System \citep[PanSTARRS;][]{2004SPIE.5489...11K} DR1}}

\newglossaryentry{lsst}{name=LSST, description={Large Synoptic Survey Telescope}, first={Large Synoptic Survey Telescope \citep[LSST;][]{2006AAS...209.8604P}}}
\newglossaryentry{ppmxl}{name=PPMXL, description={PPMXL Catalog of Positions and Proper Motions on the ICRS \citep{2010AJ....139.2440R}}}
\newglossaryentry{gaia}{name=GAIA, description={Global Astrometric Interferometer for Astrophysics \citep{2001A&A...369..339P}}, first={Global Astrometric Interferometer for Astrophysics \citep[GAIA;][]{2001A&A...369..339P}}}
\newglossaryentry{ligo}{name=LIGO, description={Laser Interferometer Gravitational Wave Observatory}, first={Laser Interferometer Gravitational Wave Observatory \citep[LIGO;][]{1992Sci...256..325A}}}
\newglossaryentry{aligo}{name=Advanced LIGO, description={Advanced LIGO}, sort=ligo2}
\newglossaryentry{lisa}{name=LISA, description={Laser Interferometer Space Antenna \citep{1994ESAJ...18..219J}}, first={Laser Interferometer Space Antenna \citep[LISA;][]{1994ESAJ...18..219J}}}
\newglossaryentry{ccd}{name=CCD,description={Charged Coupled Device}, first={charged coupled device (CCD)}, firstplural={charged coupled devices (CCDs)}}

\newcommand{\sn}[2]{SN~#1#2\xspace}

\newglossaryentry{irc}{name=IRC, text={IRC}, description={infrared catastrophe}, first={infrared catastrophe \citep[IRC;][]{1980PhDT.........1A}}}

\newglossaryentry{sn}{name=Supernova, text={SN}, plural={SNe}, description={exploding star}, nonumberlist=true, first={supernova (SN)}, firstplural={supernovae (SNe)}}
\newglossaryentry{snia}{name=Type~Ia (SN~Ia), text={SN~Ia}, description={Thermonuclear explosion of a white dwarf - spectra show no hydrogen but a strong silicon line},first={Type~Ia supernova (SN~Ia)}, firstplural={Type Ia supernovae (SNe~Ia)}, plural={SNe~Ia}, parent=sn, nonumberlist=true}
\newcommand{\sneia}{\glspl*{snia}\xspace}

\newglossaryentry{branchnormal}{name={branch-normal}, text=\textit{Branch-normal}, description={Large homogeneous class of Type Ia Supernovae, defined in \citet{1993AJ....106.2383B}}, first={\textit{Branch-normal} SNe Ia \citep{1993AJ....106.2383B}}, parent=snia}
\newglossaryentry{91t}{name={91T-like}, description={Luminous class of Type Ia supernovae similar to \sn{1991}{T} \citep{1992AJ....103.1632P}} , first={91T-like}, parent=snia}
\newglossaryentry{91bg}{name={91bg-like}, description={Faint class of Type Ia supernovae similar to \sn{1991}{bg} \citep{1992AJ....104.1543F}}, first={91bg-like}, parent=snia}
\newglossaryentry{02cx}{name={02cx-like}, description={Peculiar class of Type Ia supernovae similar to \sn{2002}{cx} \citep{2003PASP..115..453L}}, first={02cx-like \sneia\ \citep{2003PASP..115..453L}}, parent=snia}

\newglossaryentry{snibc}{name=Type~Ib/c, text={SN~Ib/c}, description={Collapse of the core of a massive star -  spectrum shows no hydrogen and no silicon line},first={Type~Ib/c supernova (SN~Ib/c)}, firstplural={Type~Ib/c supernovae (SNe~Ib/c)}, plural={SNe~Ib/c}, parent=sn}

\newglossaryentry{snib}{name=Type~Ib, text={SN~Ib}, description={Spectrum shows no hydrogen and no silicon, but helium line},first={Type Ib supernova (SN~Ib)}, firstplural={Type~Ib supernovae (SNe~Ib)}, plural={SNe~Ib}, parent=snibc}

\newglossaryentry{snic}{name=Type~Ic, text={SN~Ic}, description={Spectrum shows no hydrogen, no silicon and no helium line},first={Type~Ic supernova (SN~Ic)}, firstplural={Type~Ic supernovae (SNe~Ic)}, plural={SNe~Ic}, parent=snibc}

\newglossaryentry{snii}{name=Type~II, text={SN~II}, description={Collapse of the core of a massive star - spectrum shows strong hydrogen line},first={Type~II supernova (SN~II)}, firstplural={Type~II supernovae (SNe~II)}, plural={SNe~II}, parent=sn}

\newglossaryentry{sniib}{name=Type~IIb, text={SN~IIb}, description={Spectrum shows hydrogen and helium lines},first={Type~IIb supernova (SN~IIb)}, firstplural={Type~IIb supernovae (SNe~IIb)}, plural={SNe~IIb}, parent=snii}

\newcommand{\sniib}{\gls{sniib}}

\newglossaryentry{sniip}{name=Type~II~Plateau (Type IIP), text={SN~IIP}, description={Lightcurve shows plateau},first={Type~IIP supernova (SN~IIP)}, firstplural={Type~II Plateau supernovae \citep[SNe~IIP;][]{1979A&A....72..287B}}, plural={SNe~IIP}, parent=snii}

\newglossaryentry{sniil}{name=SN~II~Linear, text={SN~IIL}, description={Lightcurve shows no plateau, but linear decline},first={Type~IIL supernova (SN~IIL)}, firstplural={Type~II~Linear supernovae \citep[SNe~IIL;][]{1990MNRAS.244..269S}}, plural={SNe~IIL}, parent=snii}

\newglossaryentry{sniin}{name=Type II narrow-lined (Type IIn), description={Spectrum shows narrow lines},first={Type~II~narrow-lined supernova (SN IIn)}, firstplural={Type~IIn supernovae (SNe~IIn)}, plural={SNe~IIn}, parent=snii}

\newglossaryentry{snr}{name=Remnant (SNR), text=SNR, description={Remnant left visible post-explosion}, first={supernova remnant (SNR)}, firstplural={supernova remnants (SNRs)}, parent=sn}

\newglossaryentry{dtd}{name=DTD,description={delay time distribution - expected supernova rate over time after a brief outburst of starformation},first={delay time distribution (DTD)}, firstplural={delay time distributions (DTDs)}, plural=DTDs}

\newglossaryentry{hvg}{name=HVG,description={high velocity gradient - Type Ia supernovae with a fast evolution of photospheric velocity},first={high velocity group (HVG)}, firstplural={high velocity groups (HVGs)}, plural=HVGs, parent=snia}

\newglossaryentry{lvg}{name=LVG,description={low velocity gradient - Type Ia supernovae with a slow evolution of photospheric velocity},first={low velocity group (LVG)}, firstplural={low velocity groups (LVGs)}, plural=LVGs, parent=snia}

\newglossaryentry{wd}{name=white dwarf (WD), text=WD, description={White Dwarf - extremely dense stellar remnant}, first={white dwarf (WD)}}
\newglossaryentry{onemgwd}{name= Oxygen/Neon (ONe), text={ONe-WD},description={Oxygen/Neon White Dwarf}, first={oxygen/neon White Dwarf (ONe-WD)}, parent=wd}
\newglossaryentry{cowd}{name=carbon/oxygen (CO), text={CO-WD}, description={carbon/oxygen white dwarf}, first={carbon/oxygen white dwarf (CO-WD)}, firstplural = {carbon/oxygen white dwarfs (CO-WDs)}, parent=wd}

\newglossaryentry{sds}{name=SD-Scenario,description={single-degenerate scenario (single white dwarf accreting from non-degenerate companion)}, first={single-degenerate scenario (SD-scenario)}}

\newglossaryentry{dds}{name=DD-Scenario, description={double degenerate scenario (merging of two white dwarfs)}, first={double-degenerate scenario (DD-scenario)}}

\newglossaryentry{sss}{name=SSS, text={supersoft \xray\ source}, description={supersoft \xray\ source - believed to be emitted by nuclear fusion on a white dwarf's surface}}%, first={supersoft \xray\ source (SSS)}, firstplural={supersoft \xray\ sources (SSS)}}

\newglossaryentry{amcvn}{name=AM CVn, description={AM Canum Venaticorum star \citep[white dwarf accreting hydrogen poor matter from a companion star; see ][]{2005ASPC..330...27N}}}

\newglossaryentry{rlof}{name=RLOF, description={Roche Lobe Overflow (see \citet{1971ARA&A...9..183P} for a more detailed description)}, first={Roche-lobe overflow (RLOF)}}

\newglossaryentry{mchan}{name={Chandrasekhar mass~}, text={Chandrasekhar~mass}, symbol={\ensuremath{M_\textrm{Chan}}}, plural={Chandrasekhar~masses}, description={Mass when the core of a star collapses due to insufficient degeneracy pressure - for a white dwarf $\approx1.38\,M_\odot$ see \citet{1931ApJ....74...81C}}, first={Chandrasekhar~mass \citep[$M_\textrm{Chan}=1.38\,M_\odot$;][]{1931ApJ....74...81C}}, sort=mchan}

\newglossaryentry{w7}{name={W7 model},description={W7 model \citep{1984ApJ...286..644N}},first = {W7 model \citep{1984ApJ...286..644N}}}

\newglossaryentry{ew}{name=Equivalent Width, text={EW}, description={width of a rectangle that has the same area as a spectral line when taken to zero flux}, first={equivalent width (EW)}, firstplural={equivalent widths (EWs)}}
\newglossaryentry{agb}{name=AGB,description={Asymptotic Giant Branch}, first={Asymptotic Giant Branch (AGB)}}
\newglossaryentry{cmb}{name=CMB,description={Cosmic Microwave Background}}
\newglossaryentry{csm}{name=CSM,description={Circumstellar Medium}, first={circumstellar medium (CSM)}}
\newglossaryentry{csi}{name=CSI,description={Circumstellar Interaction}, first={circumstellar interaction (CSI)}}
\newglossaryentry{ism}{name=ISM,description={Interstellar Medium}, first={interstellar medium (ISM)}}
\newglossaryentry{ige}{name=IGE,description={Iron Group Element}, first={iron group element (IGE)}, firstplural={iron group elements (IGEs)}}
\newglossaryentry{epm}{name=EPM,description={Expanding Photosphere Method \citep{1974ApJ...193...27K}}, first={Expanding Photosphere Method (EPM)}}
\newglossaryentry{aic}{name=AIC,description={Accretion Induced Collapse}, first={accretion induced collapse (AIC)}}
\newglossaryentry{ime}{name=IME,description={Intermediate Mass Element}, first={intermediate mass element (IME)}, firstplural={intermediate mass elements (IMEs)}}
\newglossaryentry{h0}{name=\ensuremath{H_0},description={Hubbles constant}}
\newglossaryentry{nse}{name=NSE,description={Nuclear Statistical Equilibrium}, first={nuclear statistical equilibrium (NSE)}}
\newglossaryentry{cdm}{name=CDM,description={Cold Dark Matter}}
\newglossaryentry{grb}{name=GRB,description={Gamma Ray Burst}, first={Gamma Ray Burst (GRB)}, firstplural={Gamma Ray Bursts (GRBs)}}
\newglossaryentry{xps}{name=XPS, description={X-ray point source}, first={X-ray point source (XPS)}, firstplural={X-ray point sources (XPS)}}
\newglossaryentry{donor}{name=donor,description={non-degenerate companion in the \gls{sds}}}
\newglossaryentry{mainsequence}{name=main sequence,description={main sequence star}}
\newglossaryentry{redgiant}{name=red giant,description={red giant star}}
\newglossaryentry{mlcs}{name=MLCS,description={Multicolor Light Curve Shape method \citep[MLCS;][]{1996ApJ...473...88R}}, first={Multicolor Light-Curve Shape method \citep[MLCS;][]{1996ApJ...473...88R}}}
\newglossaryentry{rsoph}{name=RS~Ophiuci ,description={white dwarf accreting from a red giant - assumed progenitor of the \gls{sds}}, sort=rsoph}
\newglossaryentry{usco}{name=U~Scorpii,description={white dwarf accreting from a main sequence star - assumed progenitor of the \gls{sds}}, sort=usco}
\newglossaryentry{rcw86}{name=RCW~86,description={supernova remnant sometimes associated with \sn{185}{}}, sort=rcw86}
\newglossaryentry{casa}{name=Cas~A,description={Cassiopeia A supernova remnant - probably a \gls{snib} event}}
\newglossaryentry{cepheid}{name=Cepheid,description={very luminous variable star with a strong luminosity period relationship}}
\newglossaryentry{urca}{name=Urca, text=\textit{Urca}, description={process predominatly contributing to cooling in stars. The \textit{Urca} process consists of alternating electron-capture and $\beta^{-}$ decay of two nuclei pairs.},sort=urca}
\newglossaryentry{alphacen}{name=Alpha Centauri,description={one of the brightest stars in the night sky and a close binary}}
\newglossaryentry{pcygni}{name={P Cygni}, text={P Cygni},description={a hypergiant luminous blue variable with strong winds. Often referred to as a description for their line profiles showing a emission peak at the rest wavelength of the line and a blue-shifted absorption trough.}}

\newglossaryentry{teff}{name={effective temperature~}, text={effective temperature}, symbol={\ensuremath{T_\textrm{eff}}}, description={Temperature of a blackbody emitting the same total energy}, sort=teff}

\newglossaryentry{logg}{name={surface gravity~}, text={surface gravity}, symbol={\ensuremath{\textrm{log}\,g}}, description={gravity at the surface of a star}, sort=logg}
\newglossaryentry{feh}{name={metallicity~}, text={metallicity}, symbol=\textrm{[Fe/H]},description={iron abundance relative to the sun}, sort=feh}

\newglossaryentry{texp}{name={time since explosion~}, text={time since explosion}, text={time since explosion}, symbol={\ensuremath{t_{\rm exp}}},description={time since explosion (measured in days)}, sort=texp, first={time since explosion (\ensuremath{t_{\rm exp}})}}

\newglossaryentry{lmc}{name=LMC,description={Large Magellanic Cloud}, first={Large Magellanic Cloud (LMC)}, sort=lmc}
\newglossaryentry{smc}{name=SMC,description={Small Magellanic Cloud}, sort=smc}
\newglossaryentry{z}{name=\ensuremath{z},description={redshift}, sort=z}
\newcommand{\teff}{\glssymbol*{teff}}

\newglossaryentry{stats.pdf}{name=PDF, description={Probability Density Function}, first={Probability Density Function}}

\begin{document}

%   \title{Surviving companions of Cas A}
%   \title{Ruling out progenitor scenarios for the core-collapse supernova that produced Cas A}
\title{No surviving non-compact stellar companion to Cassiopeia A}
\author{Wolfgang~E.~Kerzendorf\inst{1}
  \and Tuan~Do\inst{2}
  \and Selma~E.~de~Mink\inst{3}
  \and Ylva~G\"{o}tberg\inst{3}
  \and Dan~Milisavljevic\inst{5}
  \and Emmanouil~Zapartas\inst{3}
  \and Mathieu~Renzo\inst{3}
  \and Stephen~Justham\inst{7,8}
  \and Philipp~Podsiadlowski\inst{4}
  \and Robert~A.~Fesen\inst{6}
}

\institute{European Southern Observatory, Karl-Schwarzschild-Stra{\ss}e 2, 85748 Garching bei M\"{u}nchen, Germany
  \and UCLA Galactic Center Group, Physics and Astronomy Department, UCLA, Los Angeles, CA 90095-1547, USA
  \and Anton Pannekoek Institute for Astronomy, University of Amsterdam, 1090 GE Amsterdam, The Netherlands
  \and Department of Astrophysics, University of Oxford, Keble Road, Oxford OX1 3RH, United Kingdom
  \and Department of Physics and Astronomy, Purdue University, 525 Northwestern Avenue, West Lafayette, IN 47907, USA
  \and 6127 Wilder Lab, Department of Physics \& Astronomy, Dartmouth College, Hanover, NH 03755, USA
  \and  School of Astronomy \& Space Science, University of the Chinese Academy of Sciences, 19A Yuquan Road, Beijing 100049, China
  \and National Astronomical Observatories, Chinese Academy of Sciences, 20A Datun Road, Beijing 100012, China}

  \abstract
   {
   Massive stars in binaries can give rise to extreme phenomena such as X-ray binaries and gravitational wave sources after one or both stars end their lives as core-collapse supernovae. Stars in close orbit around a stellar or compact companion are expected to explode as ``stripped-envelope supernovae'', showing no (Type Ib/c) or little (Type IIb) signs of hydrogen in the spectra, because hydrogen-rich progenitors are too large to fit. The physical processes responsible for the stripping process and the fate of the companion are still very poorly understood.

Aiming to find new clues, we investigate Cas~A, which is a very young ($\sim$340 \,yr) and near ($\sim$3.4\,kpc) remnant of a core-collapse supernova. Cas~A has been subject to several searches for possible companions, all unsuccessfully. We present new measurements of the proper motions and photometry of stars in the vicinity based on deep HST ACS/WFC and WFC3-IR data. We identify stellar sources that are close enough in projection but using their proper motions we show that none are compatible with being at the location of center at the time of explosion, in agreement with earlier findings.

Our photometric measurements allow us to place much deeper (order-of-magnitude) upper limits on the brightness of possible undetected companions. We systematically compare them with model predictions for a wide variety of scenarios. We can confidently rule out the presence of any stellar companion of any reasonable mass and age (main sequence, pre main sequence or stripped) ruling out what many considered to be likely evolutionary scenarios for \gls{sniib}.

More exotic scenarios that predict the presence of a compact companion (white dwarf, neutron star or black hole) are still possible as well as scenarios where the progenitor of Cas~A was single at the moment of explosion (either because it was truly single, or resulted from a binary that was disrupted, or from a binary merger). The presence of a compact companion would imply that Cas~A is of interest to study exotic outcomes of binary evolution. The single-at-death solution would still require fine-tuning of the process that removed most of the envelope through a mass-loss mechanism yet to be identified. We discuss how future constraints from Gaia and even deeper photometric studies may help to place further constraints.

}

\maketitle
%
%________________________________________________________________
\section{Introduction}

The deaths of massive stars mark the birth of neutron stars (NS) or black holes \citep[BH; e.g.][]{2002RvMP...74.1015W}.  They are accompanied by a bright \gls{sn} in the case of a successful explosion, during which the remaining outer layers of the progenitor star are ejected, inserting momentum, energy and newly synthesized heavy elements into the surrounding interstellar medium. There are many supernova types attributed to the core-collapse mechanism. Type II (P)lateau supernovae are clearly identified as the collapse of massive stars. The light curve plateau is explained by a recombination wave racing through the expanding hydrogen envelope. Type II (L)inear supernova decline much faster suggesting much smaller envelopes. Type IIb only show minimal traces of hydrogen. Type Ib and Type Ic show no hydrogen with the latter not even showing helium lines. Despite extensive work, many mysteries remain concerning the progenitor evolution, explosion mechanism and especially the possibly important role of binarity.

Various studies have shown that young massive stars are predominantly found in close binary (or higher-order multiples) systems \citep[e.g.][]{2007ApJ...670..747K, 2012MNRAS.424.1925C, 2012Sci...337..444S, 2014ApJS..213...34K, 2017A&A...598A..84A}.
Systems, where a companion is still present at the moment of the explosion and where that companion remains bound to the newly formed compact object, are progenitors for several of the exotic products of binary evolution. These include X-ray binaries, rejuvenated pulsars and, if the system also survives the death of the second star, a binary containing two neutron stars and/or black holes  \citep[e.g.][]{2006csxs.book..623T}. The latter are now known to give rise to short-gamma ray bursts, kilonovae and strong emission of gravitational waves when they coalesce \citep{2017PhRvL.119p1101A}.

Observationally, a large diversity is found in the characteristics that can be inferred from the light curves and spectra of core-collapse supernova.  We distinguish two main classes of core-collapse supernovae: one hydrogen-rich type showing evidence for the presence of an extended and massive hydrogen envelope (Type II) and one lacking or showing only minimal evidence for hydrogen  \citep[Type Ib/c and Type IIb respectively; see e.g.][for definitions]{1997ARA&A..35..309F}.  Supernovae that eventually give rise to double compact objects are expected to be of the second category. The presence of a close companion does not allow for enough space for an extended hydrogen-rich progenitor.

In this work, we focus on the intriguing subclass Type IIb, which can be considered as a transitional type between the hydrogen-rich (Type II) and the stripped-envelope supernovae (Type Ib/c).  The spectra of these supernovae initially show traces of hydrogen, but they disappear at later times, indicating that the progenitor has lost most but not all of its envelope.  Several theoretical studies proposed this subtype to originate from the partially stripped star in an interacting binary  \citep{1993Natur.364..509P,2009MNRAS.396.1699S,2011A&A...528A.131C,2017ApJ...840...10Y}. Alternatively, the progenitor could be a binary system that merged after a common envelope phase leaving only a thin layer of hydrogen \citep{1995PhR...256..173N}.   A single star origin has also been considered, where stellar winds and or eruptive mass loss episodes removed most but not all of the envelope \citep{1994ApJ...429..300W,2012A&A...538L...8G}. The theoretical models are subject to substantial uncertainties and therefore it is not well known whether and how much these channels contribute or if there are further explanations.

The most famous example of a supernova of Type IIb is SN 1993J \citep{1993ApJ...415L.103F}. Binary models explaining its characteristics of this specific supernova appeared shortly after its detection \citep{1993Natur.364..509P,1995PhR...256..173N}, but also more recent studies have been devoted to this supernova \citep{2009MNRAS.396.1699S,2011A&A...528A.131C}.  The close distance of 3.6~Mpc and the predictions by the early theoretical models motivated searches for a possible surviving companion. A detection of a putative companion was first claimed by \citet{2004Natur.427..129M} with a later more extensive analysis by \citet{2014ApJ...790...17F}.  Aside from SN 1993J, possible detections of a companion for Type IIb SN include SN 2011dh \citep{2014ApJ...793L..22F,2015MNRAS.454.2580M} and SN 2001ig \citep{2006MNRAS.369L..32R}.

The interpretation of the results for these extragalactic SNe is not trivial. Massive stars are often found in associations or clusters. This leads to a non-negligible chance for an unrelated bright star to be present at the location of the explosion. Moreover, at these extragalactic distances, even very deep HST observations can only retrieve the brightest companions, while theoretical simulations predict that companions are usually not very bright \citep{2017ApJ...842..125Z}.  These predictions appear consistent with the fact that several extensive campaigns searching for companions in relatively nearby supernovae only provided non-detections.

For example,  \citet{2016ApJ...818...75V} explored the explosion site of the Type Ic SN 1994I 20\,yr after the explosion, providing an upper limit on the mass of a possible main-sequence companion of $10\,\msun$. For the Type Ic-BL SN 2002ap observations taken 14 years after explosion provide a limit of $\lesssim 10\,\msun$ for the mass of a possible MS companion \citep{2007MNRAS.381..835C,2017ApJ...842..125Z}.

A much more promising strategy is to survey nearby young supernova remnants.  Their vicinity allows for the possible detection and characterization of companions to much deeper limits.  The companion may still be bound to the newly formed compact object. However, it is more likely to be unbound and fleeing the explosion site with a velocity similar to the linear velocity it had when it was still in orbit.  The expected velocities can be tens to hundreds of \kms, fast enough to be characterized as a runaway star \citep{1961BAN....15..265B,2001A&A...365...49H,2011MNRAS.414.3501E}, although low velocities of only a few \kms\ are expected to be more typical \citep[][]{2014ApJ...782....7D, 2018arXiv180409164R}.  Several runaway stars have been tentatively linked to supernova remnants that also host neutron stars \citep{2011ApJ...743L..22D,2013MNRAS.435..879T,2014AN....335..981T,2015MNRAS.448.3196D,2017A&A...606A..14B}.

In this work, we focus on the nearby supernova remnant Cas~A, which was likely created in an explosion that occurred about 340 yrs ago \citep{2006ApJ...636..848F}.  Detection of optical echoes of the supernova outburst and subsequent spectra \citep{2008Natur.456..617K, 2008ApJ...681L..81R} show large similarity to those of SN 1993J, strongly suggesting that Cas~A was also the result of a core-collapse supernova of the transitional subtype Type IIb. \citet{2006ApJ...640..891Y} infer a progenitor with an initial mass of 15 -- 25~\msun\ that has lost most but not all of its hydrogen envelope. The similarity with SN 1993J, and the fact that progenitors in this mass range are generally considered not to have winds strong enough to lose their envelope \citep{2003ApJ...591..288H, 2018MNRAS.475...55B}, has lead many to consider a binary scenario for the progenitor of Cas~A.

Cas~A is very suitable for a companion search because it is very nearby \citep[3.4 kpc; see ][]{1995ApJ...440..706R, 2014MNRAS.441.2996A} and very young ($\approx340$~years).  Its evolution has been monitored for decades \citep{1954ApJ...119..206B,2014ApJ...789..138P}. The remnant, emitting across X-ray, optical, and infrared wavelengths, has been mapped in three dimensions \citep{1995ApJ...440..706R,2010ApJ...725.2038D,2015Sci...347..526M}, providing extensive insight in the properties and explosion dynamics of a core-collapse supernova.
An X-ray point source has been detected by \citet{1999IAUC.7246....1T}, which is likely the remnant neutron star of the object that exploded as Cas A.  Given the known center of expansion and the position of the neutron star, \citet{2006ApJ...645..283F} infer a transverse velocity for the neutron star of $\approx 350$\,km\,s$^{-1}$.

The area around Cas A's center has been imaged extensively \citep[][and references therein]{1976ApJS...32..351K, 2006ApJ...636..848F, 2014ApJ...789..138P}.  No obvious companion candidate similar to that of SN1993J has ever been reported,  but there are a number of dimmer stars that have not been examined for proper motions.
Cas~A offers a unique opportunity to identify candidates due to the relatively precise knowledge of the center of expansion \citep[accuracy of $1^{\prime\prime}$;][]{2001AJ....122..297T}. This allows us to distinguish a companion from a foreground or background object by requiring it to have a proper motion that would place it near the center at the time of explosion.

Recently, \citet{2018MNRAS.473.1633K} carried out a search for binary
companions to Cas~A focusing on \gls{ps1dr1} data for both the
photometry and the extinction.  The \gls{ps1dr1} extinction maps
\citep{2015ApJ...810...25G} indicate a relatively low extinction of
$A_V \simeq 5$~mag and with this extinction the mass of a companion is
limited to being well under $1M_\odot$. \citet{2018MNRAS.473.1633K}
note that \citet[][]{2017MNRAS.465.3309D} suggest a much higher
extinction, and that for the maximum possible $A_V\simeq 15$~mag this would weaken the limit to $\simeq 2.5 M_\odot$.  They also note
that the HST photometry of a region encompassing companion ejection velocities of $300$\kms\ rules out companions with masses above 1~$M_\odot$ even at these very high extinctions. \citet{2018MNRAS.473.1633K} also only discusses the limits on companions in the context of isochrones for single stars.

Here we carry out a detailed analysis of a wider HST field including proper motion measurements for all the stars that might be companions and assuming the higher extinction estimates of  \citet{2017MNRAS.465.3309D}.  The \citet{2017MNRAS.465.3309D} and  \gls{ps1dr1} extinctions may be mutually consistent if the extra extinction is located in a sheet located very close to Cas~A.  We find no possible surviving companion stars and explore the implications of the resulting magnitude limits not only for the masses of single stars but also for stellar remnants (white dwarfs, neutron stars) and stars that have been stripped through binary interactions.

In Section~\ref{sec:data_analysis},  we present the datasets as well as the methods that were used to obtain photometry and astrometry from the candidate stars. The presented simple comparisons are then confronted with theoretical evolutionary scenarios in Section~\ref{sec:evol_scenario} and we conclude the paper in Section~\ref{sec:conclusion}.

\section{Data \& Analysis}
\label{sec:data_analysis}
The observations presented in this work have been compiled from WFPC-2 (F450W, F675W), ACS/WFC (F850LP) and WFC3-IR (F098M) observations (PI: Fesen; for all datasets used, see Table~\ref{tab:obs_table}). We use all datasets except the F850LP dataset to measure the SEDs of the candidate stars (none of the candidates are detected in F450W, but we use the detection limit from this filter set) and but only the WFC3-IR and ACS/WFC data for astrometry.

For this work, we have looked for surviving companions with escape velocities up to 600~\kms. This is an extremely generous range. Velocities are expected to be well below 300 \kms  \citep[e.g.][]{2011MNRAS.414.3501E}. Velocities below about 10~\kms\ may be most typical \citep{2018arXiv180409164R}.  Faster unbound companions are very rare and require more exotic scenarios \citep[cf.][]{2015MNRAS.448L...6T}.

Assuming an explosion date of 1680 \citep{2001AJ....122..297T} and a distance of 3.4~kpc \citep{1995ApJ...440..706R}, we calculate the search radius to be $\approx12\arcsec$. We detect 7 sources in the redder bands (all of them in F098M and some of them in F675W) of the observations within this radius and label them in Table~\ref{tab:photometry}.

\begin{table}
\begin{tabular}{lllll}
\toprule
    ID & Instrument & Filter &        date &   t\_exp \\
- & - & - & YYYY-MM-DD & s\\
\midrule
  9238 &      WFPC2 &   F450W & 2002-01-18 &    5600 \\
  9238 &      WFPC2 &   F675W & 2002-01-18 &    4000 \\
  9890 &        ACS &  F850LP & 2003-11-13 &    $4\times500$ \\
 10286 &        ACS &  F850LP & 2004-12-04 &    $4\times500$ \\
 12300 &       WFC3 &   F098M & 2010-10-29 &  $4\times553$ \\
 12674 &       WFC3 &   F098M & 2011-11-18 &  $4\times553$ \\
\bottomrule

\end{tabular}

\caption{HST Observations that are presented in this work. The ID corresponds to the HST GO proposal ID.}
\label{tab:obs_table}
\end{table}

\subsection{Photometry}
\begin{figure}[h!]
 \centering
 \resizebox{\hsize}{!}{\includegraphics{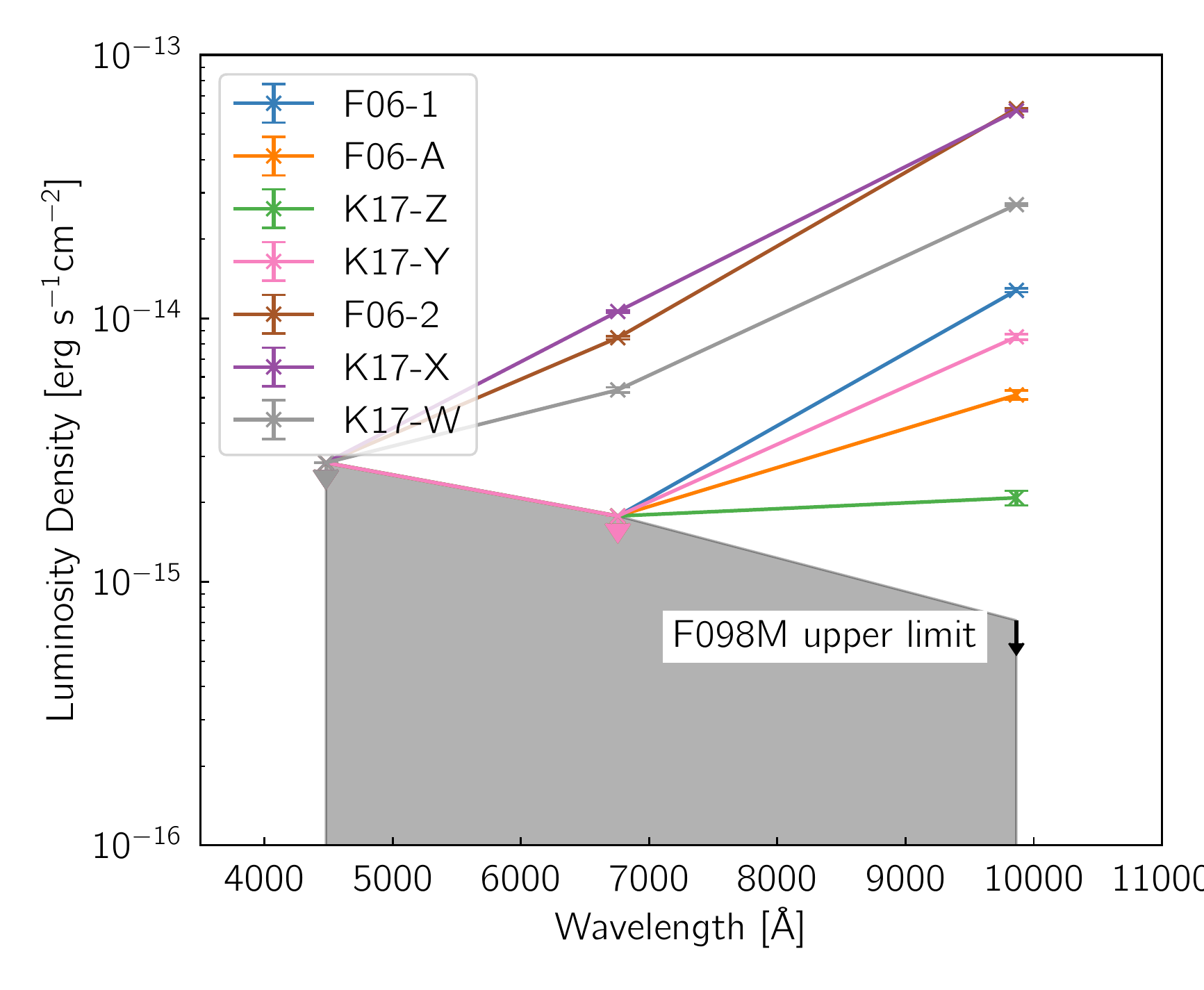}}
    \caption{Spectral energy distributions (SEDs) of the candidates constructed from the photometry measured from all three bands (F450W, F675W, F098M) including upper limits (constructed by using the magnitude of the faintest well-detected star in the entire image).}
       \label{fig:pure_seds}
\end{figure}

We use the Hubble Legacy Archive (HLA) and used their \gls{sextractor} generated data set for the photometry for our analysis for the WFPC2 and WFC3-IR images. We do not include the F850LP for the photometric measurement of this work as the separation between both chips runs through our search field and thus the photometry is not evenly measured. We used the \textit{TOTMAG} magnitude (aperture magnitude with aperture correction) supplied by the HLA (see Table~\ref{tab:photometry}). The \textit{TOTMAG} magnitude is unavailable for the F098M dataset (pipeline problem; priv. comm.  Rick White) but we reconstructed it using an aperture correction of 0.14 mag (from the handbook) on the \textit{MagAp2} magnitudes  ($0.27\arcsec$ Aperture). Finally, we use the measured photometry as well as upper limits to construct SEDs \citep[using the synthetic photometry package \gls{wsynphot};][]{kerzendorf_wolfgang_2018_1321241} that can then be further compared to theoretical models (see Figure~\ref{fig:pure_seds}).

\begin{table*}
\begin{threeparttable}
\begin{tabular}{lrrrllrr}
\toprule
 Name &     RA &     Dec & Distance &        F675W & $\sigma_{\rm F675W}$ & F098M & $\sigma_{\rm F098M}$ \\
          &    deg      &    deg      &     $\arcsec$   &mag      &       mag               &   mag    &   mag                    \\
\midrule
F06-1 & 350.8634 & 58.8131 &     4.88 &  ND\tnote{a} &          ND\tnote{a} & 21.90 &                 0.03 \\
F06-A & 350.8656 & 58.8116 &     7.63 &  ND\tnote{a} &          ND\tnote{a} & 22.96 &                 0.04 \\
K17-Z & 350.8652 & 58.8163 &     9.25 &  ND\tnote{a} &          ND\tnote{a} & 23.91 &                 0.01 \\
K17-Y & 350.8703 & 58.8126 &     9.61 &  ND\tnote{a} &          ND\tnote{a} & 22.56 &                 0.03 \\
F06-2 & 350.8632 & 58.8112 &    10.07 &      23.2009 &               0.0155 & 20.30 &                 0.04 \\
K17-X\tnote{b} & 350.8700 & 58.8161 &    11.78 &      22.9519 &               0.0122 & 20.50 &                 0.01 \\
K17-W & 350.8722 & 58.8136 &    12.18 &      23.6959 &               0.0237 & 21.36 &                 0.01 \\
\bottomrule
\end{tabular}
\begin{tablenotes}
\item[a] upper limit
\item [b] in common with \citet{2018MNRAS.473.1633K}
 \end{tablenotes}
\end{threeparttable}
\caption{Photometry of candidate stars within 600 \kms. RA and Dec taken from F098M HLA analysis. We used the labels for all stars mentioned \citet{2006ApJ...636..848F} and prefixed this with ``F06'' and create new labels (prefixed with ``K17'') for the stars that are only shown in this work, The positional astrometry from \citet{2006ApJ...636..848F} agrees with the HLA astrometry within their uncertainties.}
\label{tab:photometry}
\end{table*}

\subsection{Astrometry}

We measure the proper motions of the candidates using both the ACS/WFC F850LP and WFC3-IR F098M observations. Both the ACS/WFC F850LP (2003-11-13 -- 2004-12-04) and the WFC3-IR F098M (2010-10-29 -- 2011-11-18) data have a baseline of about one year. The ACS/WFC F850LP data were taken at the same position angle, but with an offset spatially, resulting in about 75$^{\prime\prime}$ of overlap between 2003-11-13 and 2004-12-04. The two epochs of WFC3-IR F098M data were taken at the same position angle and had approximately the same center. While ACS/WFC F850LP allow for more accurate astrometry due to finer spatial sampling, but not all stars are detected in F850LP. Thus we use both pairs of observations to measure the proper motions for all candidates.

For all observations, we download the individual FLC file from the HST archive and measured the positions of stars in each individual frame. The positions are measured using the PSF fitting program \textit{img2xym\_WFC.09x10} for ACS/WFC data and \textit{img2xym\_wfc3ir\_stdpsf} for WFC3-IR \citep[e.g.][]{
2000PASP..112.1360A, 2015ApJ...813...27H, 2017ApJ...844..167B}. These programs are designed specifically for HST data and take into account the filter and spatially dependent PSF of each instrument.

\begin{figure*}[h!]

  \includegraphics[width=0.49\linewidth]{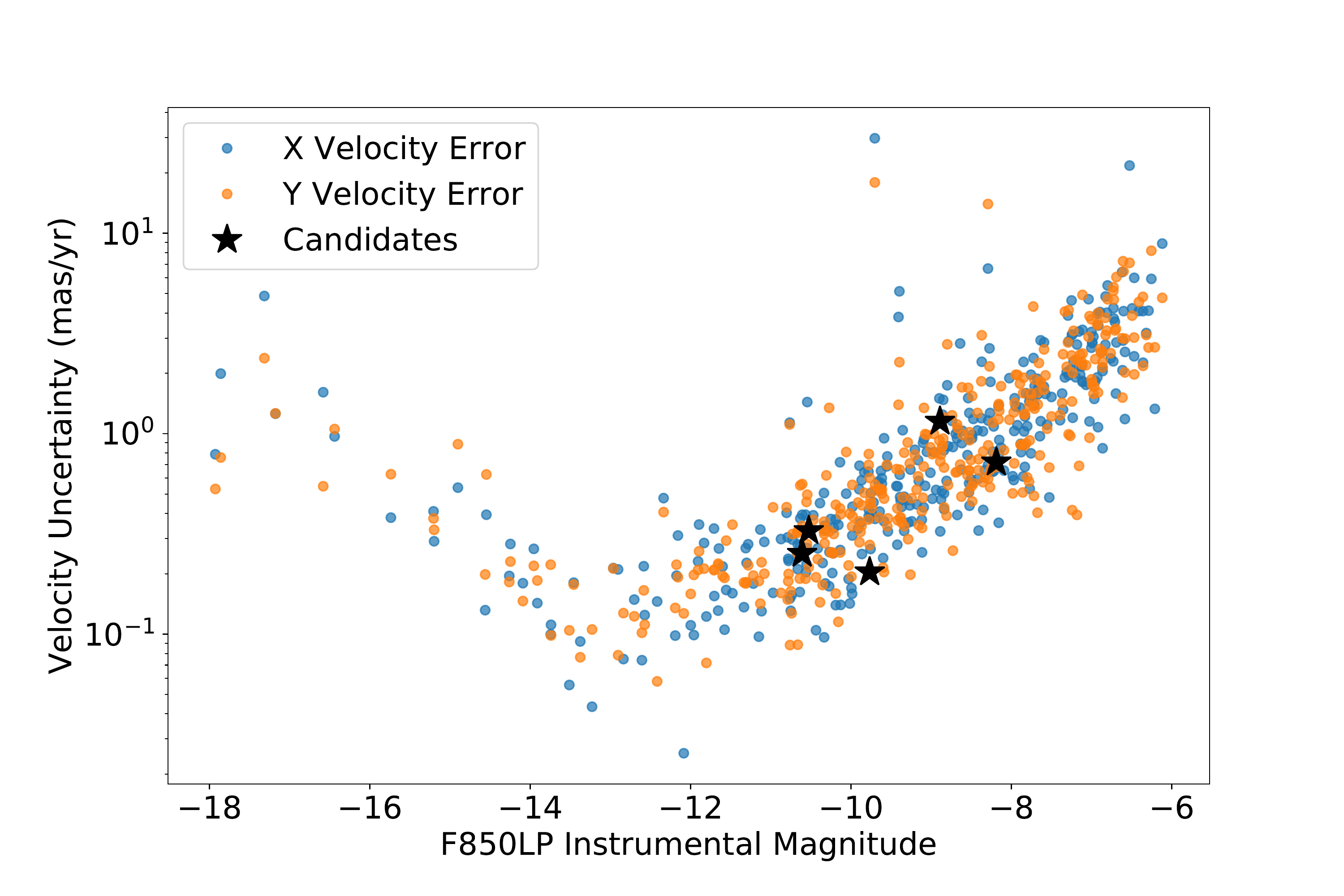}
  \includegraphics[width=0.49\linewidth]{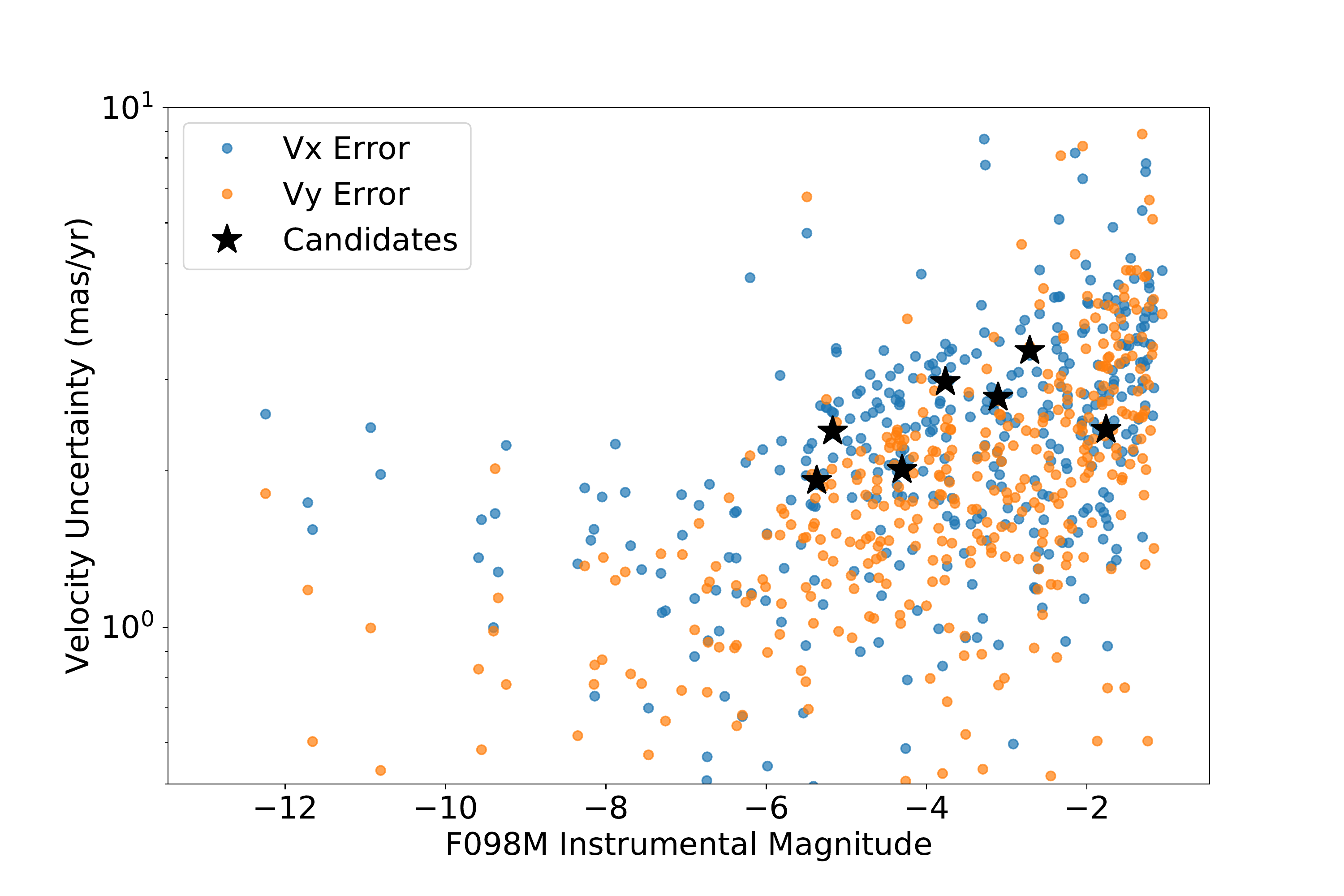}
    \caption{Proper motion uncertainty for ACS/WFC F850LP (Left) and WFC3-IR F089M (Right) in the x (blue) and y directions (orange). The candidates' companions are indicated with star symbols. The F850LP proper motion uncertainties are up to an order of magnitude better than that of F098M. The increase in uncertainty at bright magnitudes is likely due to saturation of bright sources.}
       \label{fig:vel_err}
\end{figure*}

The proper motions are measured in a reference frame created for each pair of observations using stars in common between the two epochs. We use the first epoch of observations as the reference epoch and minimized a first order linear transformation between bright stars in the two epochs. This reference frame also accounts for the optical distortion so that the final pixel positions of stars are in a distortion-free reference frame. The final position for each epoch is the average of their positions in the four frames and with an uncertainty that is the standard deviation of the positions in the four frames divided by $\sqrt{4}$. The proper motion is then the difference between the pixel positions in the first and second epoch (e.g. $PM_{x} = x_2 - x_1$, $\sigma_{PM} = \sqrt{\sigma_{x_2}^2 + \sigma_{x_1}^2}$). Figure \ref{fig:vel_err} shows the distribution of proper motion uncertainties as a function of instrumental magnitude.

\begin{figure*}[h!]
 \centering
 \includegraphics[width=\hsize]{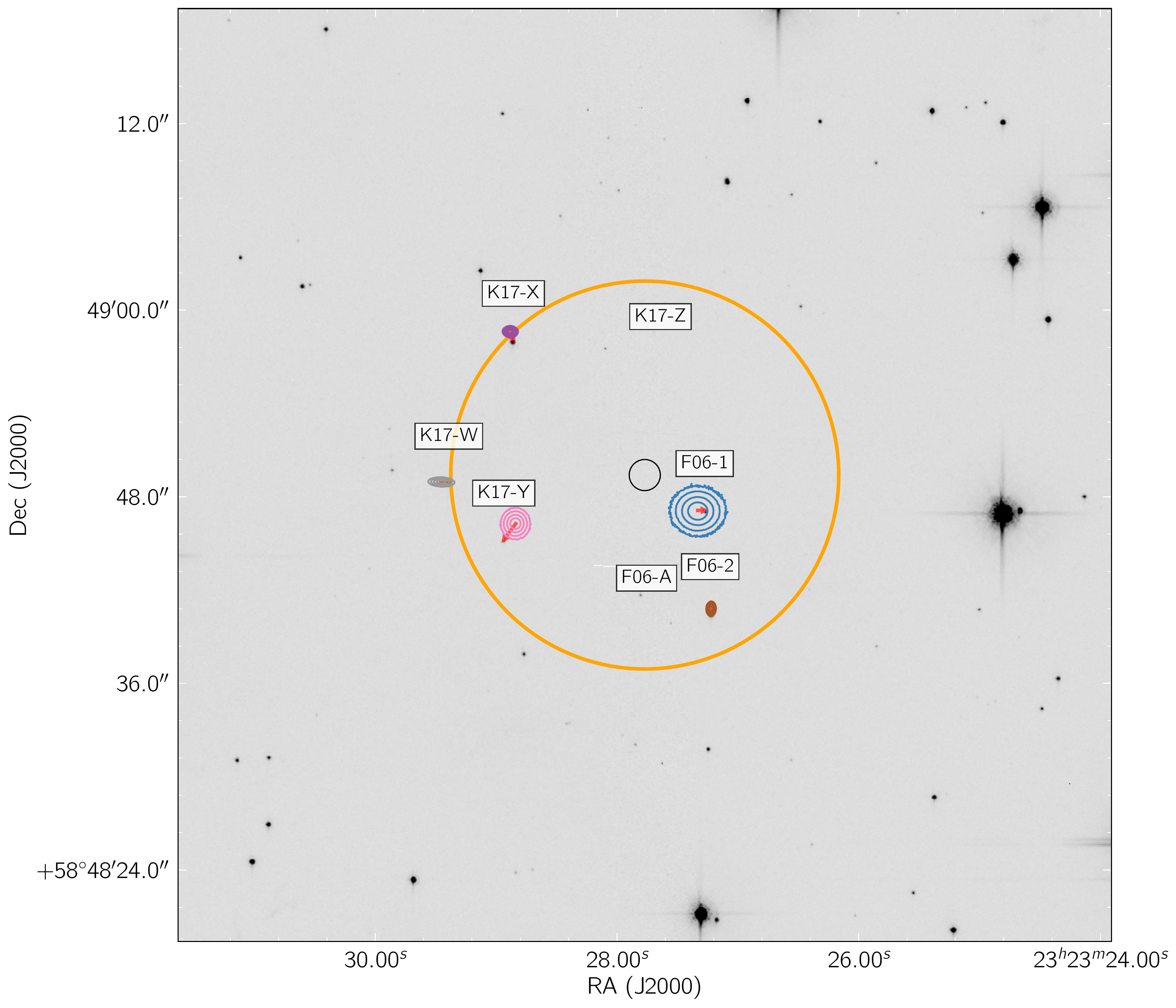}
    \caption{HST F850LP (epoch 2003-11-13) image showing the probabilities for the positions of the candidates in 1680. The center of expansion (including the uncertainty of $1\arcsec$) is depicted as a black circle. The search radius of 600~\kms around the center of expansion is shown in orange. The contours show 68\%, 95\%, 99.7\%, and 99.9937\% confidence intervals (using the same colors as in Figure~\ref{fig:pure_seds}). A red arrow marks the movement since 1680 to their current position. Some candidates are not detected in this filter and thus proper motion is not shown for them.}
       \label{fig:prop_mot_f850lp}
\end{figure*}

\begin{figure*}[h!]
 \centering
 \includegraphics[width=\hsize]{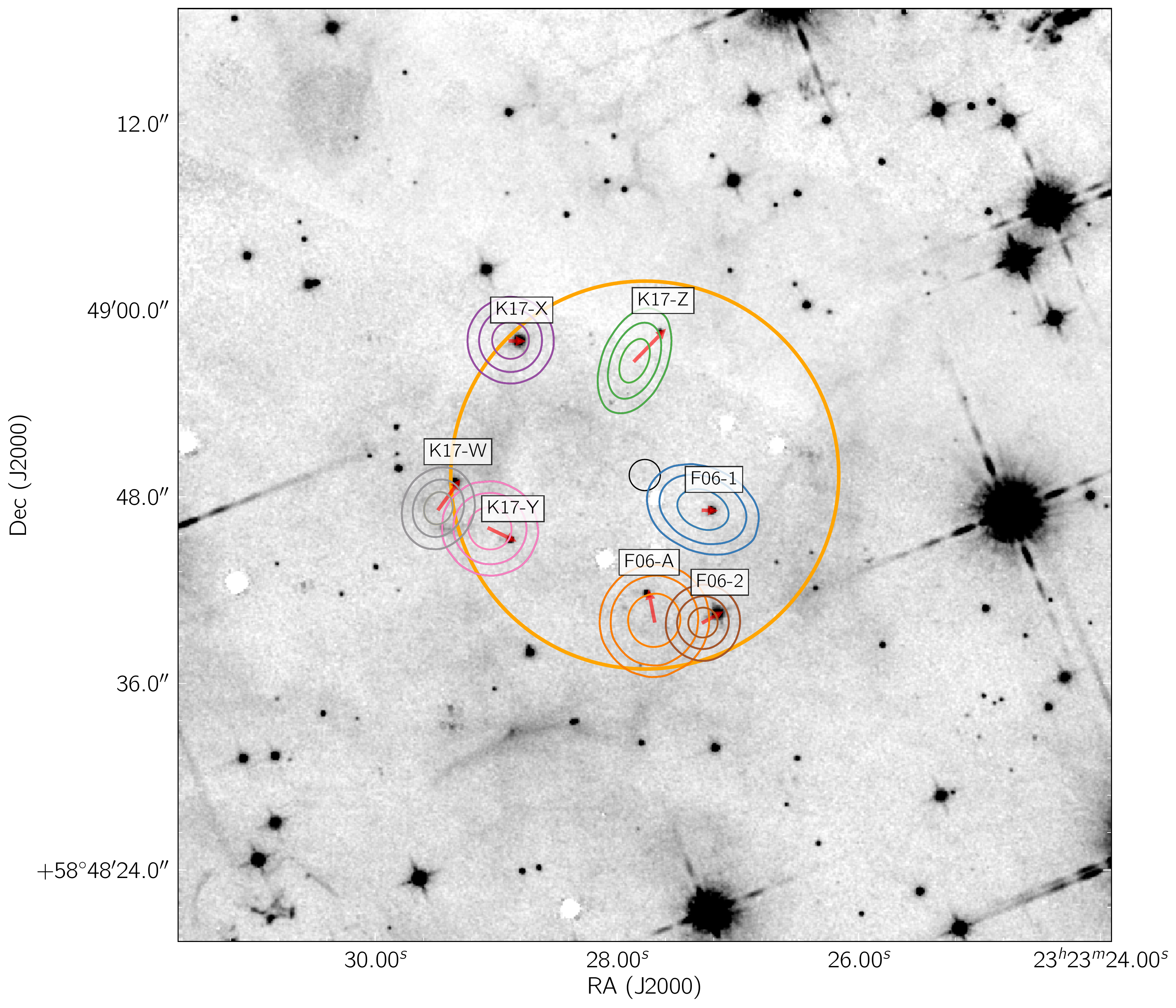}
    \caption{HST F098M image (epoch 2010-10-29) showing the probabilities for the positions of the candidates in 1680. The center of expansion (including the uncertainty of $1\arcsec$) is depicted as a black circle. The search radius of 600~\kms around the center of expansion is shown in orange. The contours show 68\%, 95\%, and 99.7\% confidence intervals (using the same colors as in Figure~\ref{fig:pure_seds}). A red arrow marks the movement since 1680 to their current position. }
       \label{fig:prop_mot_f098m}
\end{figure*}

We then filtered the resulting table of matched stars in both epochs by only allowing sources that had been detected in at least 2 frames per epoch and positional uncertainties of less than 0.5~px ($\approx 25$~mas for F850LP and $\approx 60$~mas for F098M) in both x and y positions in all frames. We found the uncertainty of the position of the stars at the time of explosion of Cas~A (assuming 1680) by performing a Monte Carlo experiment on the proper motions in pixel space and calculating the positions to 1680 (using $10^6$ samples). We then transformed these Monte Carlo pixel positions using the HLA WCS to RA and Dec. These results are shown in Figure~\ref{fig:prop_mot_f850lp}\&\ref{fig:prop_mot_f098m} for the two filters. The transformation assumes a 50 mas plate scale for ACS/WFC F850LP and 121 mas for WFC3-IR F098M. The final proper motion measurements are listed in Table~\ref{tab:astrometry}.

We compute the proper motions for the two instruments separately, rather than combine for several reasons. First, switching instruments can incur a large systematic uncertainty due to differences in optical distortion, PSF variations, etc. Second, the ACS/WFC F850LP astrometry is at least a factor of three more precise than the WFC3-IR F098M. Even with the increased time baseline from combining F850LP with F098M, the proper motions will not significantly improve. We, therefore, use F850LP proper motions for all sources except for those that are only detected in F098M, such as F06-A and K17-Z.

\begin{table*}
\footnotesize
\begin{threeparttable}
\begin{tabular}{lrrrrrrrrrr}
\toprule
Name &       RA &     Dec & F098M $\mu_\alpha$ & F098M $\sigma_{\mu_\alpha}$ & F098M $\mu_\delta$ & F098M $\sigma_{\mu_\delta}$ & F850LP $\mu_\alpha$ & F850LP $\sigma_{\mu_\alpha}$ & F850LP $\mu_\delta$ & F850LP $\sigma_{\mu_\delta}$ \\
 &       deg   &  deg        &    mas~yr$^{-1}$                &          mas~yr$^{-1}$                   &        mas~yr$^{-1}$            &              mas~yr$^{-1}$               &         mas~yr$^{-1}$            &            mas~yr$^{-1}$                  &          mas~yr$^{-1}$           &                 mas~yr$^{-1}$             \\
\midrule
F06-1 & 350.8634 & 58.8131 &               -1.5 &                         6.7 &               -0.5 &                         5.2 &                -1.5 &                          2.2 &                -0.1 &                          2.2 \\
F06-A & 350.8656 & 58.8116 &                1.6 &                         6.7 &                4.9 &                         6.8 &                 ND\tnote{a} &                          ND\tnote{a} &                 ND\tnote{a} &                          ND\tnote{a} \\
K17-Z & 350.8652 & 58.8163 &               -4.9 &                         3.9 &                5.3 &                         6.0 &                 ND\tnote{a} &                          ND\tnote{a} &                 ND\tnote{a} &                          ND\tnote{a} \\
K17-Y & 350.8703 & 58.8126 &               -3.8 &                         5.5 &               -2.5 &                         5.5 &                 2.3 &                          1.4 &                -3.5 &                          1.5 \\
F06-2 & 350.8632 & 58.8112 &               -2.7 &                         3.8 &                1.5 &                         3.8 &                 0.1 &                          0.4 &                -1.2 &                          0.7 \\
K17-X & 350.8700 & 58.8161 &               -1.5 &                         4.7 &               -0.5 &                         4.8 &                -0.7 &                          0.7 &                -1.9 &                          0.5 \\
K17-W & 350.8722 & 58.8136 &               -3.2 &                         3.7 &                4.6 &                         4.4 &                -1.6 &                          1.0 &                -0.4 &                          0.4 \\
\bottomrule
\end{tabular}

\begin{tablenotes}
\item[a] Non-detection
\end{tablenotes}
\end{threeparttable}
\caption{Astrometry of candidate stars within 600 \kms. RA and Dec taken from F098M HLA analysis.}
\label{tab:astrometry}
\end{table*}

\section{Results}

We find that there are no candidates with proper motions that could place them at the center of expansion at the time of explosion (Figures \ref{fig:prop_mot_f850lp} and \ref{fig:prop_mot_f098m}). ACS/WFC F850LP provides the strongest constraints for all sources that are detected in that filter (F06-1, FK17-Y, F06-2, K17-X, K17-W). WFC3-IR F098M proper motions are able to eliminate the remaining two sources (F06-A and K17-Z) as coming from the center of the explosion. The proper motions of sources that are detected in both instruments are statistically consistent. The closest candidate progenitor is F06-1 at 4.9$^{\prime\prime}$. A coincidence with the center of expansion in 1680 is ruled out with more than $4\sigma$ confidence. The uncertainty in the center of expansion would need to be much larger than the current value (1$^{\prime\prime}$) for this source to be a viable progenitor to Cas~A.

This suggests that any viable candidate needs to be dimmer than the photometric
limit given by these observations (see Table~\ref{tab:photometry}). The photometric limits were derived by presenting the magnitude of the faintest star found in the HLA catalogue that has a quality flag of 0 (point-source, with trusted photometry, not crowded). This results in very conservative upper limits of $2.8\times10^{-15}, 1.8\times10^{-15}, 7.1\times10^{-16} \textrm{erg}~\textrm{s}^{-1}~\textrm{cm}^{-2}$\ in F450W, F675W, F098M respectively.

\section{Evolutionary Scenarios}
\label{sec:evol_scenario}

\gls{sniib} progenitors are stars that have been stripped of most, but not
all, of their H-rich envelopes with a ($\approx 0.01 - 0.5~\msun$ H-rich thin envelope remaining;\citealt{1993Natur.364..509P, 2010ApJ...725..940Y, 2011A&A...528A.131C, 2017ApJ...840...10Y}). Depending on the amount of hydrogen
left, they can have extended envelopes (and appear as supergiants such
as the progenitors of SN 1993J and SN 2011dh) or be relatively compact
 \citep[][]{2010ApJ...711L..40C,2017ApJ...840...10Y}. Extended and compact SNe IIb's can be
distinguished during the early supernova phase, where the lightcurve
of extended SNe IIb shows a short first peak, typically lasting
several days and being powered by the release of the energy that has
been deposited by the supernova shock in the extended envelope; this
peak is missing for compact progenitors. The light echo observations
of Cas A do not allow us to distinguish between these two possibilities,
but the evolutionary history will generally be very
different. Extended progenitors require stable Roche-lobe overflow
(RLOF) in a relatively wide binary where the donor star may still be
filling its Roche lobe at the time of the explosion or has become
detached when the envelope mass has become sufficiently small and the
donor starts to shrink. The mass in the extended envelope will
typically be several 0.1\,$M_\odot$ \citep[][]{1993Natur.364..509P}. Compact progenitors can be the result of either stable RLOF in
a relatively close binary \citep[see][]{2017ApJ...840...10Y} or the outcome of a
common-envelope phase where the envelope has been ejected during a
spiral-in phase \citep{1995PhR...256..173N}. In the latter case, the system will generally be
rather compact, and the progenitor only have a very small H-rich
envelope after the common-envelope (CE) phase. Whether the system still shows hydrogen
lines in the supernova also depends on the mass loss from the post-CE
compact star, which will depend on the metallicity of the object \citep[also see][]{2017ApJ...840...10Y}.  Another difference between extended and compact progenitors is that the former generally have to be the
primaries of the binaries as mass transfer from a massive secondary to
a compact star (the remnant of the first supernova) is expected to be
dynamically unstable\footnote{If the first supernova forms a
relatively massive black hole, this mass-transfer phase could
potentially be stable \citep{2017MNRAS.471.4256V}.}.  Therefore, for extended progenitors, one
generally expects a stellar companion at the time of the supernova and
that the companion is relatively massive (so that the mass-transfer
phase is stable), as is the case for SN 1993J and SN 2011dh. In
contrast, if a compact progenitor is formed through common-envelope
evolution, it could be either the primary or the secondary of the
binary, and the companion could be quite low in mass: it just has to
be massive enough to release enough orbital energy in the spiral-in
phase to be able to eject the common envelope. Note that, if the
compact progenitor is the original secondary of the binary, the
companion star might be a compact star; depending on the detailed
history, it could be either a white dwarf, a neutron star, or a black
hole.

We assume a distance to Cas~A of 3.4~kpc \citep[with an uncertainty of about 10\%;][corroborated by \citealt{2014MNRAS.441.2996A}]{1995ApJ...440..706R} for the comparison with theoretical evolutionary scenarios. 

\begin{figure}[h!]
\centering
\includegraphics[width=\hsize]{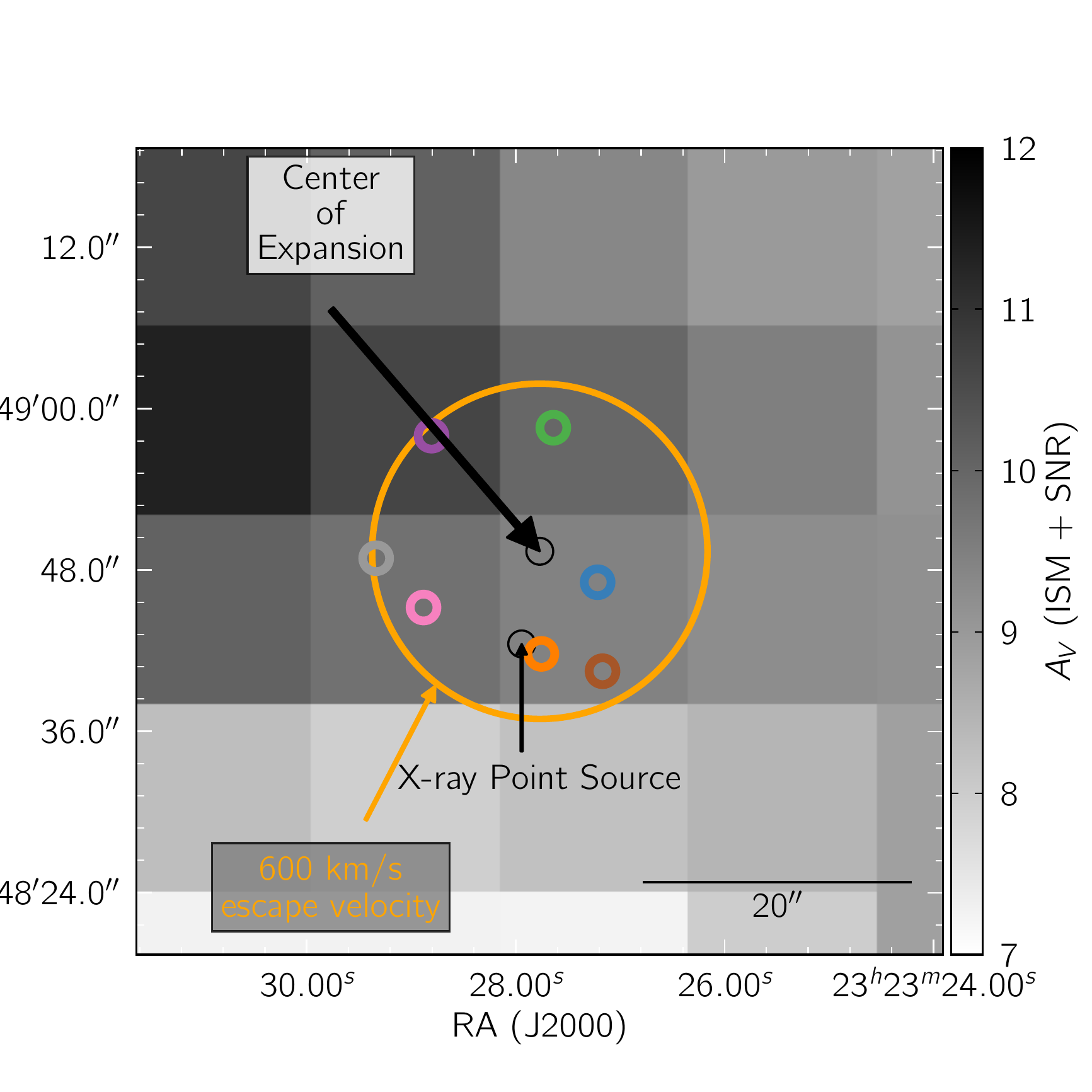}
\caption{Extinction map \citep[using the data from][]{2017MNRAS.465.3309D} assuming both the extinction from the remnant and the ISM. We have added the search radius
of 600~\kms, the center, the x-ray point source and the candidates (using the same colors as in Figure~\ref{fig:pure_seds})}
\label{fig:extinction_map}
\end{figure}

Figure~\ref{fig:extinction_map} shows the extinction map presented in \citet{2017MNRAS.465.3309D}. The center of the remnant is only covered by four extinction measurements. We choose the highest extinction value of those ($A_V=10.6$\,mag) for comparing theoretical candidate models with the photometry. The extinction might be patchy in the remnant's center and that the companion could be hidden behind a dust cloud. Thus, in addition to the measured extinction ($A_V=10.6$\,mag), we also consider a very conservative $A_V=15$\,mag in our comparisons of measured SEDs with theoretical ones. Dust emission \citep{2017MNRAS.465.3309D} is used to obtain the result of $A_V=10.6$\,mag. Thus the result might be an overestimate as the method cannot distinguish between foreground and background dust. We discuss this possibility by measuring the extinction towards stars in Appendix~\ref{sec:neighbourhood}. For this work, however, we stick with the conservative extinction estimate as this allows for stronger conclusions.

For our comparison, we only consider companion models consistent with the data if their flux is below our detection limit as none of the detected stars are companions of Cas A due to their proper motion. When referencing quantitative estimates from \citet{2017ApJ...842..125Z} in the next sections the reader should note: 1) the study includes all stripped core collapse supernovae (not only \gls{sniib}). 2) there are unquantified model uncertainties and the given numbers should be used as an estimate.

\subsection{Main Sequence Companion}

A large fraction, if not the majority of SNe stripped of their hydrogen envelope, is expected to have a main sequence companion at the moment of explosion \citep[][keeping in mind that the latter looked at all stripped core-collapse supernovae]{2014A&A...563A..83C, 2017ApJ...842..125Z}.
In Figure~\ref{fig:sed_model_ms_compare}, we compare various main sequence companion models with the measured SEDs of the companion candidates. We obtain synthetic SEDs for main-sequence stars using the temperatures and radii given in \citet{2007hsaa.book.....Z} and interpolating the BOSZ-grid \citep{2017AJ....153..234B} using \textsc{StarKit} \citep[][]{wolfgang_kerzendorf_2015_28016} and assuming the extinction and distance to Cas~A described earlier.

The data clearly rule out any companions similar to those found near other \gls{sniib} sites. Specifically, the massive hot star ($\teff\approx 24000$) found by \citet[][see also \citealt{2014ApJ...790...17F}]{2004Natur.427..129M} at the position of \sn{1993}{J} post-explosion (within 0.3~pc of the supernova site) is excluded as a companion model for Cas~A. This measurement is also in tension with the results from \citet{2017ApJ...842..125Z} which suggests that the distribution would peak around companions with $9 - 10\msun$. 

\subsection{Stripped-star Companion}

For completeness, it is worth considering the presence of a stripped-star companion, i.e. a companion that has lost its hydrogen-rich envelope as a result of a prior interaction. This type of companion is not expected to be common according to binary population synthesis (less than 5\%), but not impossible either \citep[][]{1994A&A...290..119P, 2017ApJ...842..125Z, 2018arXiv180409164R}.

We use the spectral model grids of stars stripped in binaries through stable Roche-lobe overflow presented in \citet[][see also \citealt{2017A&A...608A..11G}]{2018A&A...615A..78G}. These models were computed with the non-LTE radiative transfer code \gls{cmfgen}. We use models with solar metallicity.
% (0.3 up to 1.9 \msun for the stripped star, corresponding to initial masses 2 up to $7.4~\msun$)
We probed a grid of initial masses of the progenitor of the stripped star ($M_\textrm{init}= 2 - 18.2\msun$), which corresponds to the following masses of the stripped stars: $M_\textrm{stripped}= 0.3 - 6.7\msun$. The present data would have detected almost any stripped companion (unless for very extreme values of extinction; see Fig.~\ref{fig:sed_model_stripped_compare}), which might rule out any viable scenario for a surviving companion that was stripped pre-explosion.

\subsection{White Dwarf Companion}

A further unlikely, but certainly not impossible, scenario predicts the presence of a white dwarf companion \citep[at most a few percent for all stripped \glspl{sn}; ][]{2017ApJ...842..125Z}. This scenario has already been proposed for the formation of eccentric binaries consisting of a white dwarf and a neutron star, similar to the observed systems PSR B2303+46 and PSR J1141-6545 \citep[e.g.][]{1999MNRAS.309...26P,2000A&A...355..236T, 2006MNRAS.372..715C}.

Figure~\ref{fig:sed_model_wd_compare} shows that our limits cannot exclude a $1.2\msun$ CO white dwarf in the case of assumed high extinction \citep[corresponding to a 7\msun\ initial mass; see Figure 8 in ][]{2009ApJ...693..355W} with the temperature and radius given by \citet[][\teff=80000~K; radius=4300~km]{2011A&A...531L..19T}.

This means that the following scenario is still allowed, where Cas A was the result of the explosion of the secondary in a relatively low mass binary system.  The primary star, i.e. the initially most massive star of the system, had to be at least almost $8\msun$. After mass loss through Roche-lobe overflow, it was destined to become a white dwarf. The secondary, in this scenario, needs to gain a substantial fraction of the mass of the envelope of the primary to become massive enough to explode. This would imply that Cas~A is one of the late core-collapse supernova as discussed in \citet{2017A&A...601A..29Z}, exploding with a delay of about 50 -- 200 Myr after formation.

In principle, there would be two testable predictions that follow from this scenario. Deeper searches should eventually be able to detect the white dwarf. It should not be much older than about 200 Myr and therefore still be relatively hot ($\gtrapprox 15000$\,K). A second test would be to accurately characterize the co-moving surrounding stellar population which can be identified with Gaia. The expectation in this scenario is that this population has an age of 50 -- 200 Myr.

\subsection{Neutron Star or Black Hole companion}

\begin{figure*}[h!]

  \begin{subfigure}[t]{.47\textwidth}
   \centering
   \includegraphics[width=\textwidth]{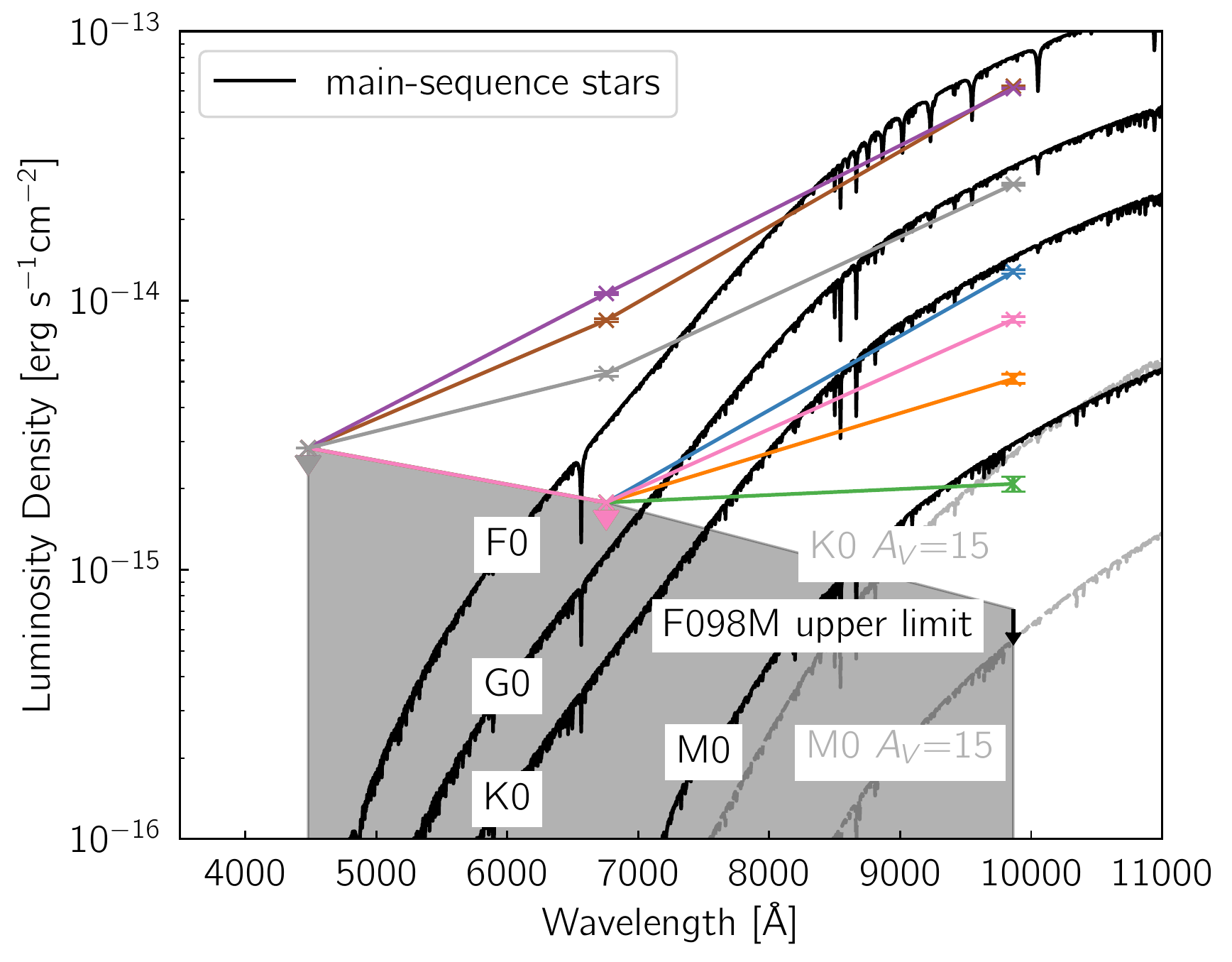}
  \caption{Synthetic SEDs (default extinction estimate in black; extreme extinction in gray) using temperatures and radii of main-sequence stars using the given the distance and extinction to Cas A compared to the measured SEDs of companion candidates (in color).}
  \label{fig:sed_model_ms_compare}
  \end{subfigure}
  \medskip
  \begin{subfigure}[t]{.47\textwidth}
   \centering
   \includegraphics[width=\textwidth]{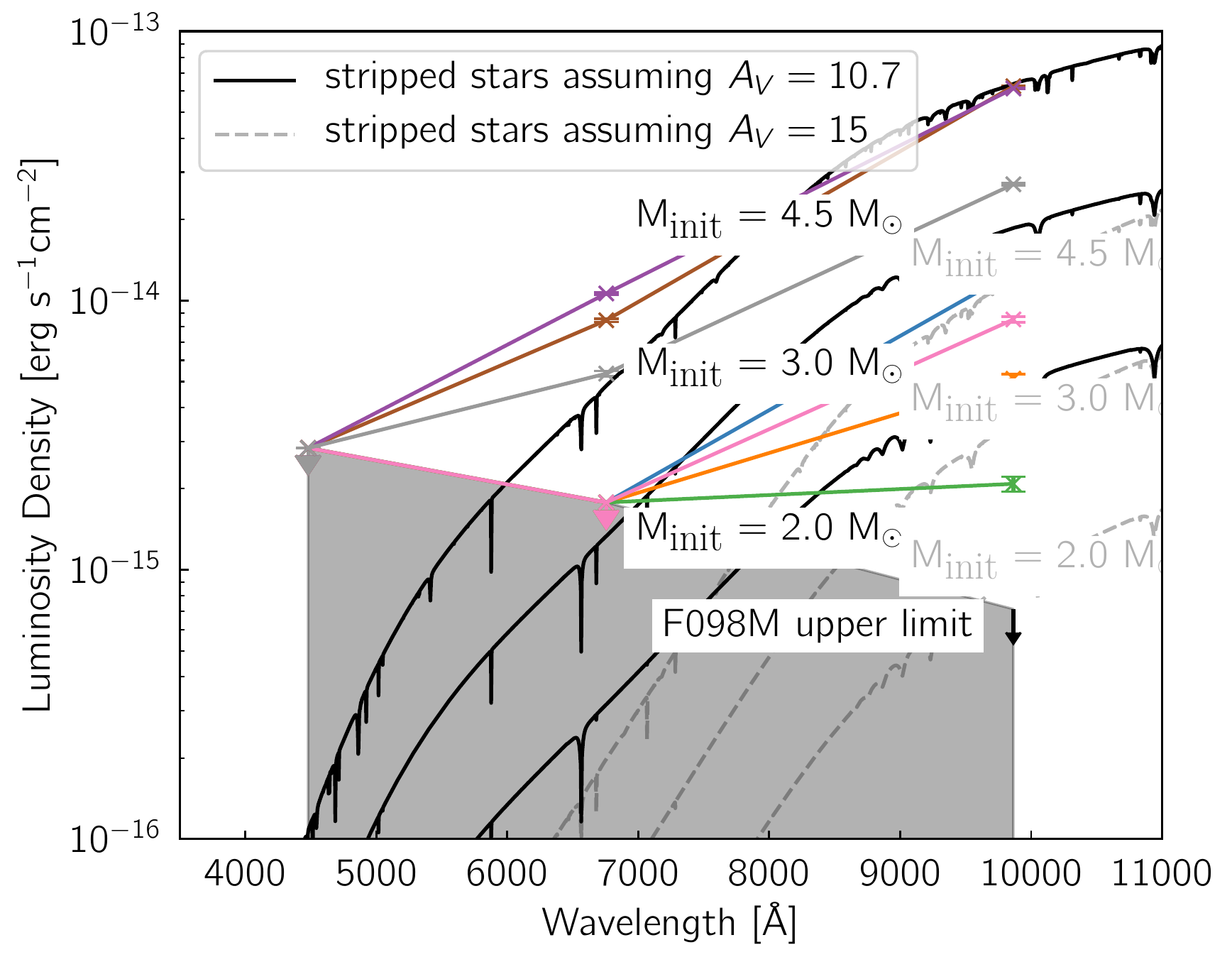}
   \caption{Synthetic SEDs (default extinction estimate in black; extreme extinction in gray) of stripped stars of 0.3 up to 1.0 \msun, corresponding to initial masses 2 up to 4.5 \msun\ using the given the distance and extinction of Cas A compared with the measured SEDs of companion candidates (in color).}
  \label{fig:sed_model_stripped_compare}
  \end{subfigure}
  \medskip
  \begin{subfigure}[t]{.47\textwidth}
   \centering
   \includegraphics[width=\textwidth]{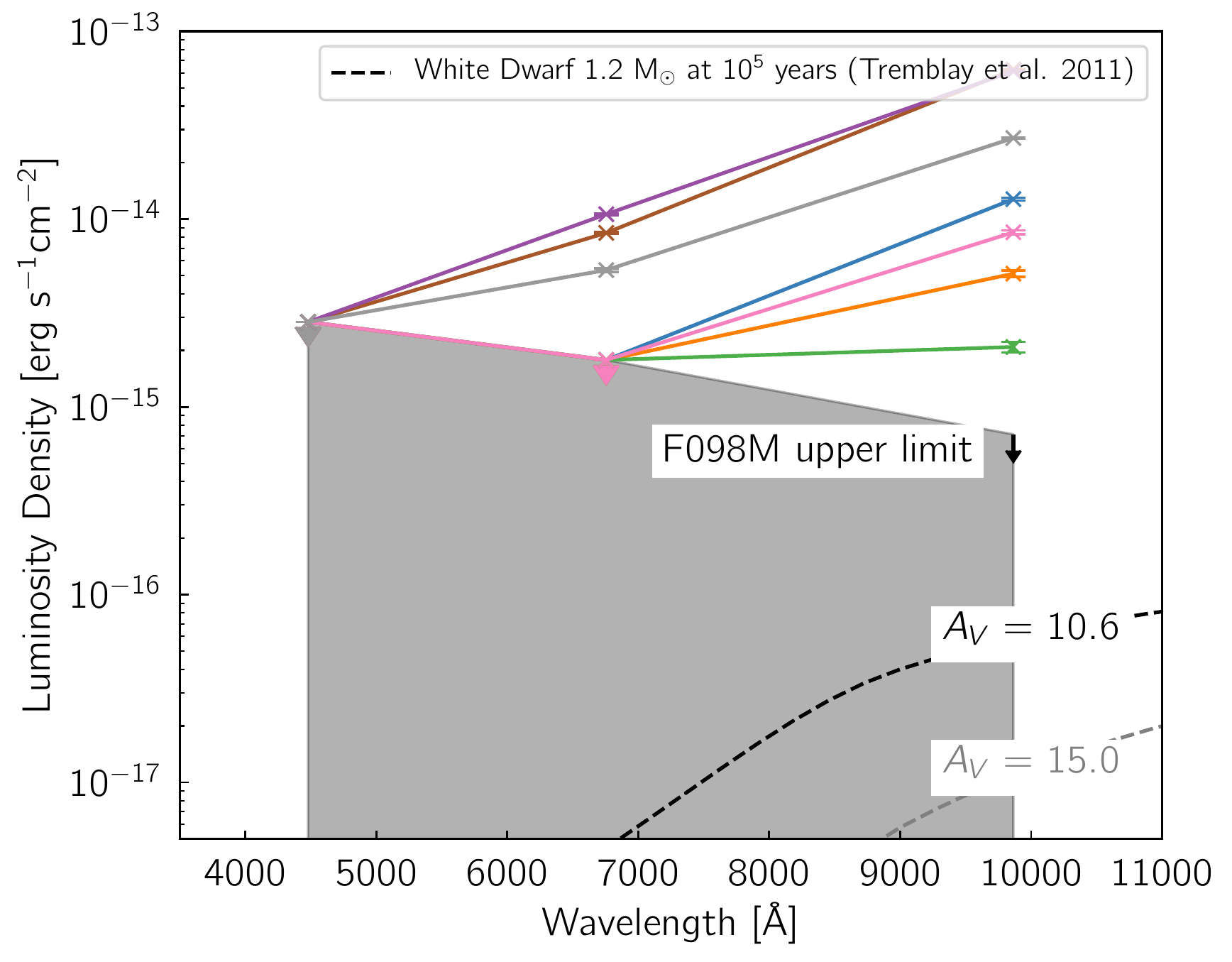}
  \caption{Synthetic SEDs (default extinction estimate in black; extreme extinction in gray) for a 1.2~\msun\ white dwarf at \num{100000} years post explosion given the distance and extinction of Cas A compared with the measured SEDs of companion candidates (in color).}
  \label{fig:sed_model_wd_compare}
  \end{subfigure}\hfill
  \begin{subfigure}[t]{.47\textwidth}
   \centering
   \includegraphics[width=\textwidth]{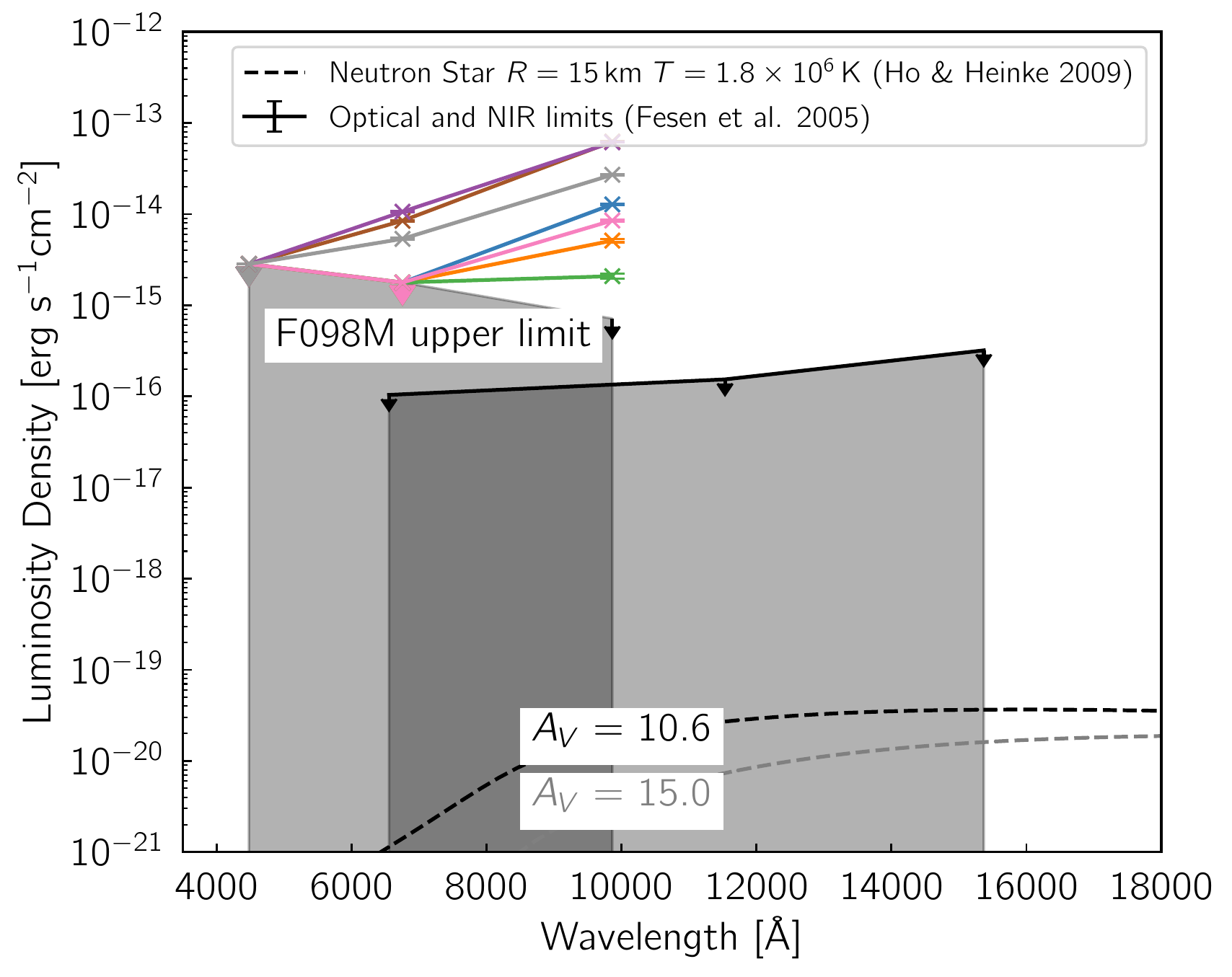}
  \caption{Neutron star SED (default extinction estimate in black; extreme extinction in gray) using the parameters from \citet{2009Natur.462...71H} and using the given the distance and extinction of Cas A compared with the measured SEDs of companion candidates (in color).}
  \label{fig:sed_model_ns_compare}
  \end{subfigure}

  %    \caption{Model comparison for various main sequence companions post explosion given the distance and extinction of Cas A \citep[using temperatures and radii from ][]{2007hsaa.book.....Z} }
   %      \label{fig:sed_model_ms_compare}
      \caption{SEDs of the candidates (using the same colors as in Figure~\ref{fig:prop_mot_f850lp}\&\ref{fig:prop_mot_f098m}) constructed from the photometry measured from all three observations including upper limits. The gray shaded regions are below the detection limit of the presented photometry. This is the only area where a surviving star would not be detected from this analysis.}
  
         %\label{fig:pure_seds}
  \end{figure*}

The rate for NS/BH companions is even smaller than for white dwarf companions \citep[$\lessapprox 1\%$;][]{2017ApJ...842..125Z} because the binary system needs to survive the disruption from the prior SN kick. \citet{1999IAUC.7246....1T} identified an \gls{xps} near the center of expansion and suggest that this is the remnant neutron star of Cas A. Very deep imaging in the optical and infrared bands \citep{2006ApJ...636..848F} have not revealed this point source (see Figure~\ref{fig:sed_model_ns_compare}). This would suggest that we will likely also miss any other older neutron star in the field with our observations. It can not be excluded that the firstborn neutron star formed with an unusually high birth kick. For example, PSRs B2011+38 and B2224+64 have inferred 2D speeds of about $1600$~\kms\ \citep{2005MNRAS.360..974H}, but no further \gls{xps} in the wider region have been identified with the remnant.

Another interesting hypothesis is that the visible \gls{xps} was formed by a previous supernova explosion in the system, while the compact object remaining from the explosion that gave rise to the Cas~A remnant has not been detected yet (e.g. a black hole). This hypothesis may not be very likely, but is still highly intriguing, because it would make Cas~A a direct progenitor of a possible binary consisting of two neutron stars or a neutron star and a black hole, if the system remained bound. This, in turn, would make it of potential interest as a gravitational wave progenitor, although we stress that other scenarios are more likely \citep[compact companions only make up 5\% of the distribution][]{2017ApJ...842..125Z}.

This hypothesis might potentially be constrained by understanding the observed cooling of the object \citep[e.g.,][]{2010ApJ...719L.167H,2013ApJ...777...22E}.  The object is consistent with being a neutron star \citep[][]{2009Natur.462...71H}, and interpretations typically reasonably assume that the object is only a few centuries old. However, if the XPS is a neutron star which was responsible for stripping the SN progenitor, then accretion may have re-heated an older object.   A related question is whether a suitable phase of accretion onto the neutron star, and perhaps also the effect of being close to the supernova, might leave behind a subsurface thermal structure which could naturally explain the unexpectedly rapid recent surface cooling of the object \citep[][]{2010ApJ...719L.167H,2013ApJ...777...22E}.

%==

\subsection{No companion}
There are three options for the Cas~A progenitor being a single star at the time of explosion. The star evolved as a single star, or it merged with a companion before its death, or it was a member of a binary system that was disrupted by a prior SN.

Single star evolution for Type IIb is possible through wind mass-loss of very massive stars \citep[see the review by][]{2007ARA&A..45..177C}. However, uncertainties in the wind mass loss rates of red-supergiants might allow for thin pre-explosion hydrogen envelopes also for initial masses below $\sim20\,\msun$ \citep[][]{2010ApJ...717L..62Y, 2017A&A...603A.118R}. The neighbourhood of Cas A contains stars that are consistent with being above $30\msun$ assuming a distance of 3.4~kpc and an appropriate extinction (see Appendix~\ref{sec:neighbourhood}). This suggests that Cas A also might have been such a massive star and thus the hypothesis of Cas A being a single star cannot be easily ruled out. However, \citet{2006ApJ...640..891Y} argue that a single star which was massive enough to eject the majority of its own envelope through winds would not naturally explain the abundance structure of the remnant.  This is primarily based on the high-velocity, nitrogen-rich and hydrogen-poor material in the remnant, which they consider a robust constraint. They also compare the $^{44}$Ti and $^{56}$Ni abundances of the remnant to their explosion models; for these, \citet{2006ApJ...640..891Y} also suggest that the evidence is more easily consistent with binary-stripped progenitors, but admit that those constraints are uncertain and model-dependent. This scenario may be further constrained by carefully age dating the nearby stars for which Gaia finds similar proper motions and radial velocities. In this scenario, we expect the surrounding population to be not much older than about 20~Myr.

The second scenario suggests a merger before explosion. \citet[][]{1995PhR...256..173N} describe the possibility of such a merger scenario for \sn{1993}{J} \citep[9 years before the binary companion was identified in ][]{2004Natur.427..129M}. They argue that a possible evolution starts with two stars with a $q<<1$ in which the primary forms a He core and expands to fill the Roche lobe. This leads to the formation of a common envelope and a subsequent spiral in of the secondary. If this spiral-in deposits enough of the orbital energy into the common envelope this might unbind it leaving it with a single star with a thin hydrogen envelope and no companion. This scenario is also consistent with the observations presented in this work.

Finding no companion is also consistent with a scenario in which the Cas~A progenitor was ejected from a disrupted binary system. The disruption occurred due to a prior explosion of the initially more massive star of the system that evolved first. In this case, Cas~A progenitor may have experienced mass exchange in the binary system before being ejected.

\section{Conclusion}
\label{sec:conclusion}

\begin{table*}[ht!]
  \begin{center}
\setlength\extrarowheight{5pt}
\begin{threeparttable}
\begin{tabular}{lcc}
\toprule
Companion Type &  A$_V$=10.6 mag& A$_V$=15 mag\\
\midrule
Main Sequence companion	 & allowed below M0 & allowed below K5\\ \hline
Stripped Stars & \multicolumn{2}{c}{not allowed}\\ \hline
White Dwarfs & \multicolumn{2}{c}{allowed}\\ \hline
Single Star & \multicolumn{2}{c}{\makecell{Stars with $>30$\msun\ exist \\ in the neighbourhood}}\\ \hline
Disrupted binary & \multicolumn{2}{c}{allowed}\\ \hline
Pre-explosion Merger & \multicolumn{2}{c}{allowed}\\ \hline
Neutron Star & \multicolumn{2}{c}{allowed}\\ \hline
Black hole & \multicolumn{2}{c}{allowed}\\ \hline
\end{tabular}
%\begin{tablenotes}
%\item[a] upper limit
%\end{tablenotes}
\end{threeparttable}
\caption{Description of compatibility of progenitor scenarios with the current data using two estimates of extinction.}
\label{tab:overview}
\end{center}
\end{table*}

We present the deepest proper motion study for a surviving companion in Cas A and explore the implications this has on several evolutionary scenarios (for an overview see Table~\ref{tab:overview}). \citet{2018MNRAS.473.1633K} has done a wider but shallower study of the stars near the Cas~A center of expansion (with our candidate K17-X and \#1 in common). He assumes a much lower extinction  \citep[$A_V\approx4$\,mag compared to our assumption of $A_V=10.6$\,mag as suggested by ][]{2017MNRAS.465.3309D} but arrives at the same conclusion as this work -- namely no companion is found for Cas~A. One might argue that the large amount of dust suggested by \citet{2017MNRAS.465.3309D} lies behind the supernova and that the extinction to Cas A is much lower (see also Appendix~\ref{sec:neighbourhood}). If this were confirmed this dataset might also rule out young and hot white dwarfs as a companion.

The only possibilities left for Cas~A's progenitor scenario are a compact remnant (white dwarf, neutron star, or blackhole), a binary merger, an ejected star from a disrupted binary system or a single star (see Table~\ref{tab:overview}). This is in contrast to the reported find of luminous companions for other \gls{sniib} such as \sn{1993}{J} (\citealt{2004Natur.427..129M}; but see also \citet{2014ApJ...790...17F}), possibly for \sn{2001}{ig} \citep{2006MNRAS.369L..32R}, and \sn{2011}{dh} \citep{2014ApJ...793L..22F}. However, for \sn{2011}{dh} the detection of a companion is questioned by \citet{2015MNRAS.454.2580M}.

The most likely predicted scenario for a Type IIb supernova suggests the presence of a main sequence companion. This appears to be ruled out for Cas A. In Section~\ref{sec:evol_scenario}, we have suggested several possible tests to further investigate the remaining possible scenarios.

The possibility that Cas~A was single at death either requires fine tuning to explain why the star ended its life with a small amount of hydrogen, or it argues in favor of late-time mass loss mechanism that are inefficient in removing the full envelope. Some mechanisms for late-time mass loss have been proposed in the literature but it is not clear why they leave a thin layer of hydrogen.
The possibility that the companion may be a white dwarf might make Cas~A of interest in light of known eccentric binaries consisting of an older white dwarf and a younger neutron star.
The possibility of a second neutron star or black hole as a companion is certainly the most intriguing one. In fact, we cannot rule out that the neutron star that is detected in the remnant is originating from the first explosion in the system, rather than the second explosion that gave rise to the remnant of Cas~A. This possibility makes Cas~A of interest for understanding the formation of binary neutron stars and binaries containing one neutron star and one black hole. The system still needs to remain bound at the second explosion. If bound, the system is a potential future gravitational wave source. We do however stress that while this may be the most intriguing possibility it is probably the least likely. In particular, the offset of the detected X-ray source from the center of expansion seems hard to explain in this scenario.

The result presented in this paper gives rise to two hypotheses when thinking about the class of \gls{sniib}. If \gls{sniib} arise from only one specific scenario then Cas~A is a typical \gls{sniib}. This suggests that no \gls{sniib} have luminous surviving companions post-explosion and the reported finds are coincidental alignments. Secondly, Cas~A is different to other \gls{sniib} like SN{1993}{J} and there are multiple progenitor scenarios that lead to these semi-stripped supernovae. Theory then predicts (see Section~\ref{sec:evol_scenario}) that there are more \gls{sniib} with companions than without. We believe that falsifying either hypothesis can likely only be done using statistical methods on all \gls{sniib} explosion sites.

\section{Acknowledgements}

W.~E.~Kerzendorf was supported by an ESO Fellowship. S.~E.~de Mink acknowledges support by H2020 ERC StG 2016 BinCosmos. We acknowledge the Princeton Center for Theoretical Science and the organizers of the workshop ``CSI: Princeton -- A Definitive Investigation of the Core-Collapse Supernova Cassiopeia A'' (supported by Adam Burrows).  This research benefited from discussions at the Kavli Institute for Theoretical Physics, which is supported in part by the National Science Foundation under Grant No. NSF PHY11-25915.

We thank Christopher Kochanek for an extensive review of the manuscript and feedback. We would like to thank Monika~Petr-Gotzens, Sabine~Moehler, Marina~Rekjuba, Jason~Spyromilio, Schuyler~Van~Dyk, Justyn Maund, and Stephen Smartt for constructive discussions and help with the manuscript.  We thank Chris Fryer for stimulating and enjoyable discussions. We thank Jay Anderson for providing openly accessible tools for the proper motion analysis. We thank the two anonymous referees for their valuable feedback that improved the paper. 

Based on observations made with the NASA/ESA Hubble Space Telescope, and obtained from the Hubble Legacy Archive, which is a collaboration between the Space Telescope Science Institute (STScI/NASA), the Space Telescope European Coordinating Facility (ST-ECF/ESA) and the Canadian Astronomy Data Centre (CADC/NRC/CSA).

The Pan-STARRS1 Surveys (PS1) and the PS1 public science archive have been made possible through contributions by the Institute for Astronomy, the University of Hawaii, the Pan-STARRS Project Office, the Max-Planck Society and its participating institutes, the Max Planck Institute for Astronomy, Heidelberg and the Max Planck Institute for Extraterrestrial Physics, Garching, The Johns Hopkins University, Durham University, the University of Edinburgh, the Queen's University Belfast, the Harvard-Smithsonian Center for Astrophysics, the Las Cumbres Observatory Global Telescope Network Incorporated, the National Central University of Taiwan, the Space Telescope Science Institute, the National Aeronautics and Space Administration under Grant No. NNX08AR22G issued through the Planetary Science Division of the NASA Science Mission Directorate, the National Science Foundation Grant No. AST-1238877, the University of Maryland, Eotvos Lorand University (ELTE), the Los Alamos National Laboratory, and the Gordon and Betty Moore Foundation.

\bibliographystyle{apj}
\bibliography{wekerzendorf}

\begin{appendix}

\section{Astrometry}

\begin{figure*}[h!]
\begin{subfigure}[t]{.47\textwidth}
 \centering
	 \includegraphics[width=\textwidth]{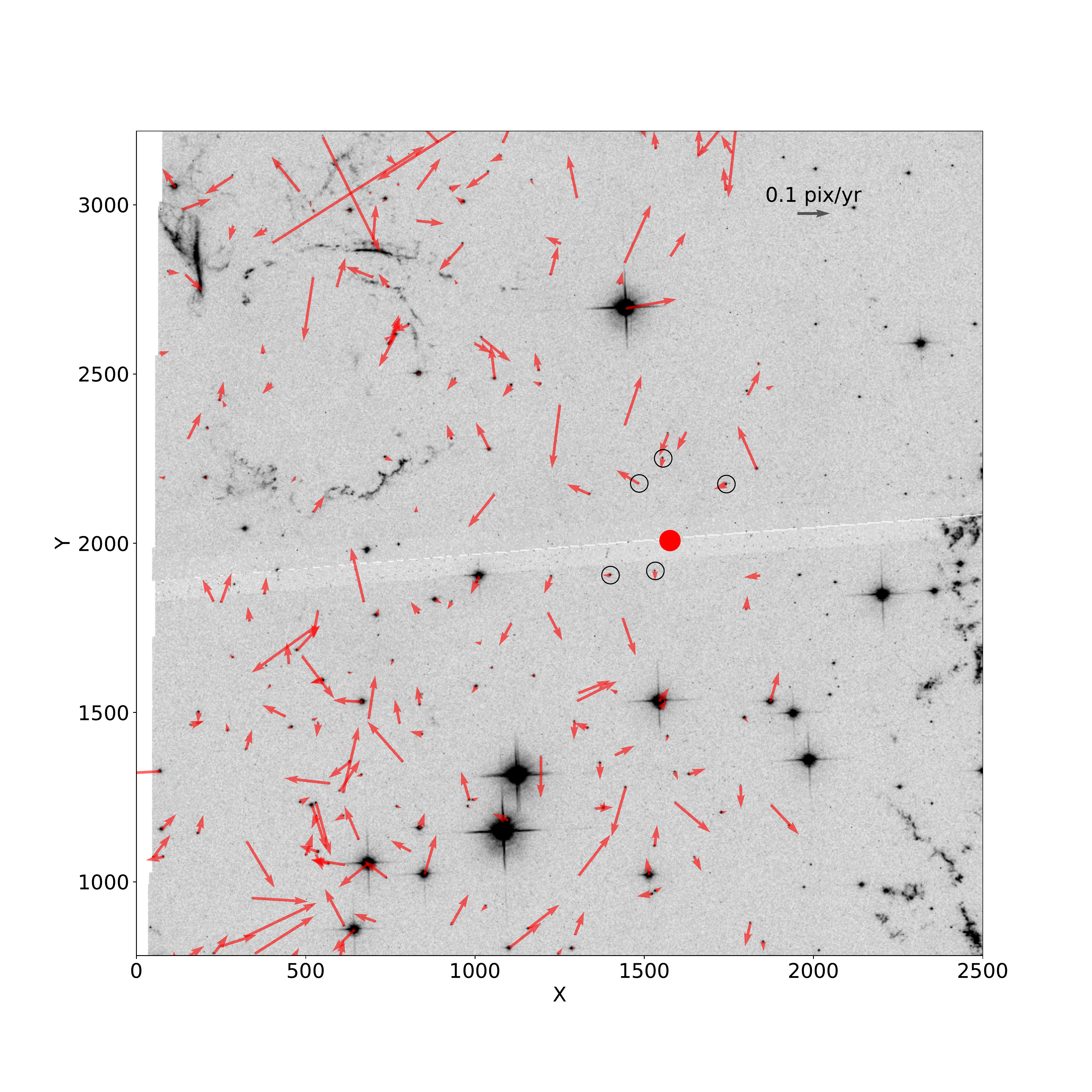}
         \caption{The proper motions of stars as measured using ACS F850LP astrometry. The background image is from 2003
         and is oriented at a position angle of 87.6 deg.
         Proper motions are available only for stars that overlap between the 2003 and 2004 pointings. The five candidate companion sources (black circles) and center of expansion (red dot) are labeled. }
       \label{fig:f850l_quiver}
 \end{subfigure}\hfill
 \begin{subfigure}[t]{.47\textwidth}
 \centering
	 \includegraphics[width=\textwidth]{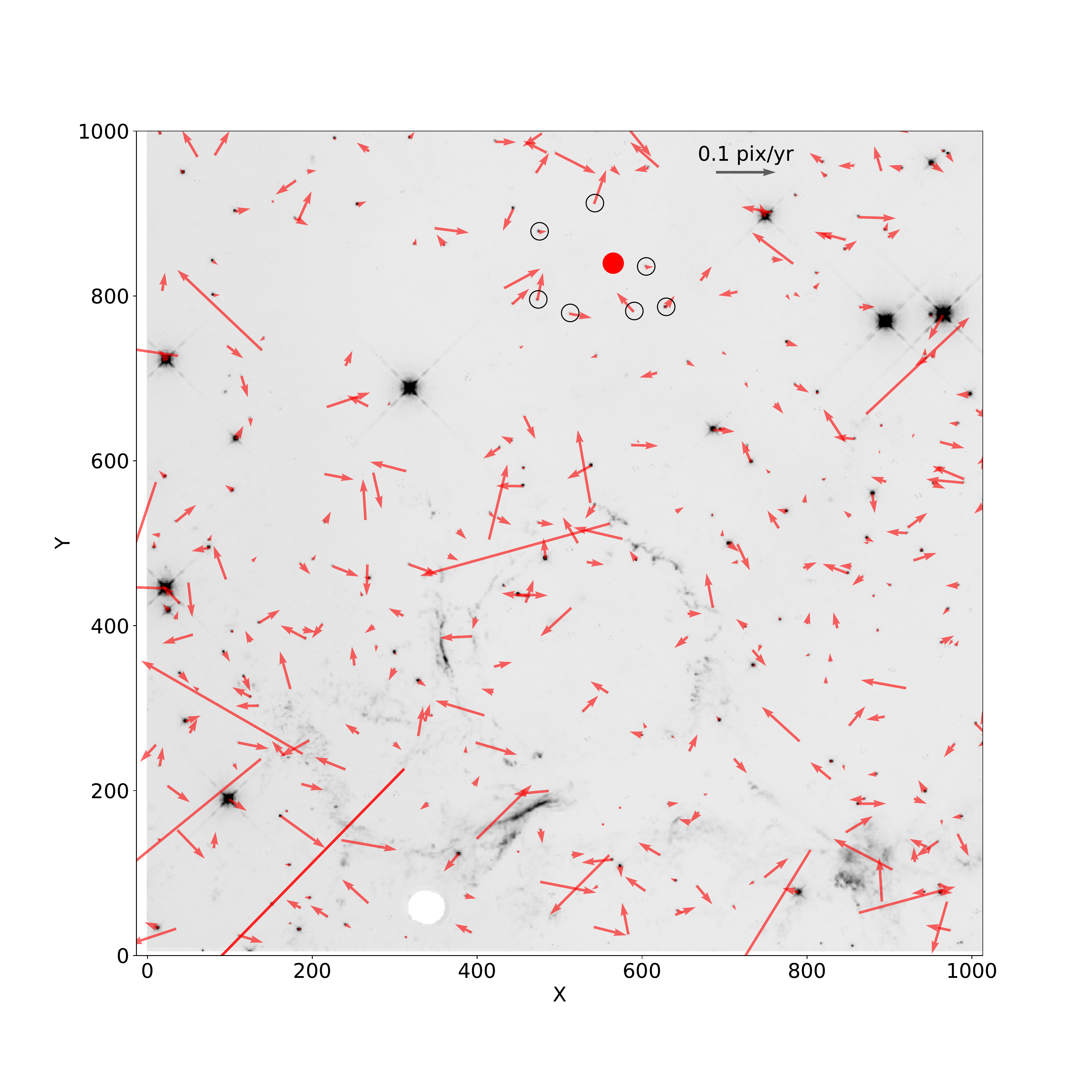}
         \caption{The proper motions of stars as measured using WFC3-IR F098M astrometry from 2010 to 2011. The background image is from 2010 and
         is oriented at a position angle of -23.3 deg.
         The five candidate companion sources (black circles) and center of expansion (red dot) are labeled. }
       \label{fig:f098m_quiver}
        \end{subfigure}
\caption{Overview of the proper motion measurements for the two datasets from ACS F850LP \& WFC3-IR F098M}
\end{figure*}

In order to determine the robustness of the astrometry, we also examine the proper motions of sources in the field outside of our search radius. In ACS/WFC F850LP, the overlap between the two epochs is about $75^{\prime\prime}\times202^{\prime\prime}$, while the WFC3-IR F098M observations have about complete overlap $123^{\prime\prime}\times136^{\prime\prime}$. By examining the proper motion of sources across the field of view, we can estimate the systematics of with some sources (such as local optical distortion). Figures \ref{fig:f850l_quiver} and \ref{fig:f098m_quiver} show the proper motion vectors in the reference frame of the instrument. There are no significant local grouping of proper motions.

\section{The neighbourhood of Cassiopeia A}
\label{sec:neighbourhood}

\begin{figure}[h!]
 \centering
	 \includegraphics[width=\hsize]{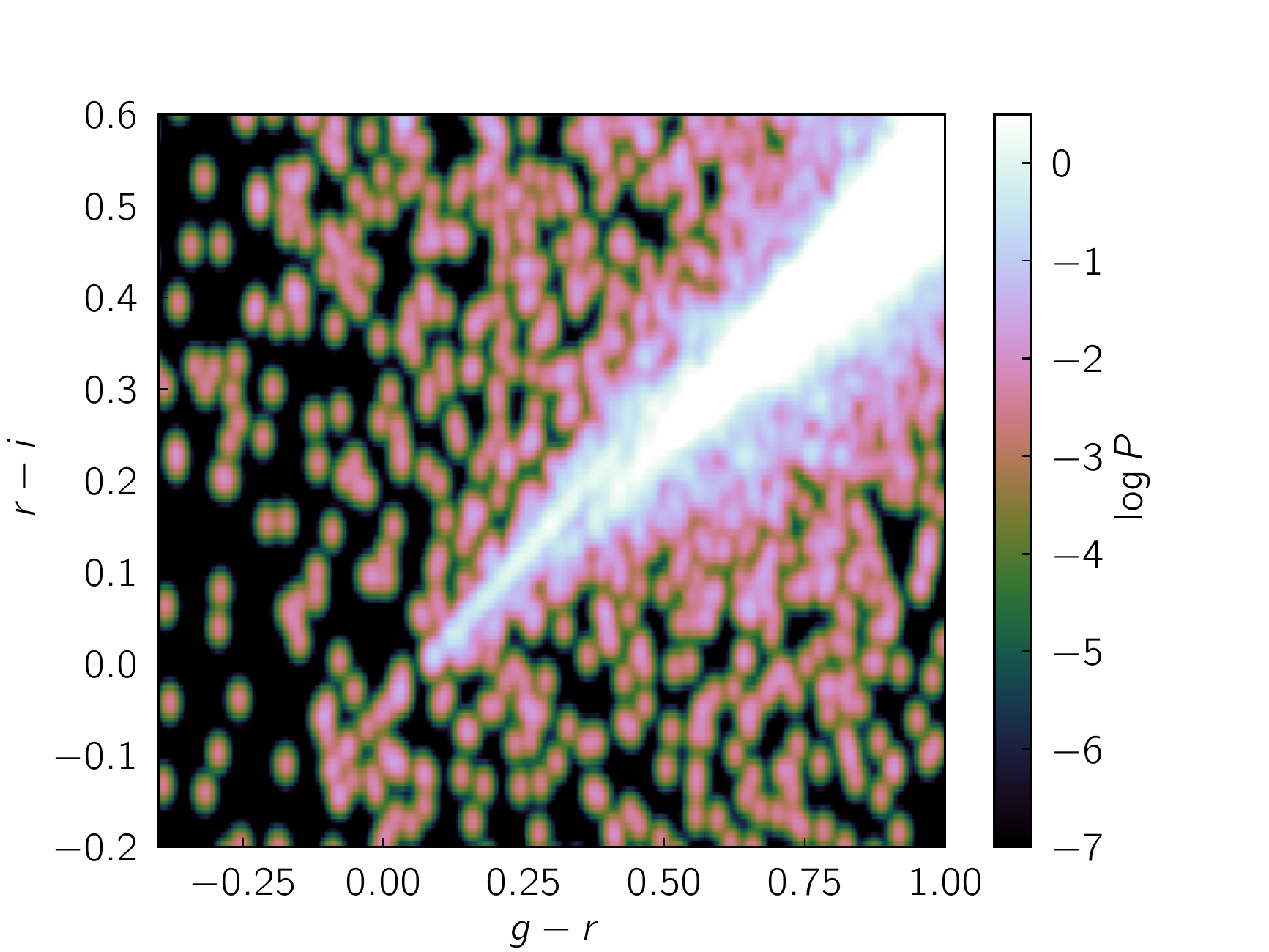}
         \caption{Gaussian kernel density estimate with \gls{scikit-learn} using a bandwidth of 0.1 of $g-r$ and $r-i$ photometry of approximately$\num{18000}$ stars with 50\arcmin of Cas A.}
       \label{fig:panstarrs_cc}
\end{figure}

The progenitor system that resulted in the supernova Cas A can be studied by looking at the surviving members of the birth cloud. Assuming that the progenitor is less than 50 Myr and a drift velocity of 5~\kms we arrive at a current size of the birthcloud of 250~pc which corresponds to 250\arcmin\ at 3.4~kpc. We acquired \gls{panstarrs} photometry within a $50\arcmin$ radius of Cas A, then reduced this dataset of $\approx\num{320000}$ stars by only selecting stars with $g-r<1$ and $r-i<1$ resulting in $\approx\num{18000}$ stars. Figure~\ref{fig:panstarrs_cc} shows the density estimate in color-color space of these stars (using a Gaussian kernel with a bandwidth of 0.1) which reveals the existence of an offset young population at the blue end of the stellar locus.

These approximate 300 stars might originate from the same cloud as the progenitor of Cas A. We use isochrones to estimate their age and properties. For \textsc{parsec} isochrones \citep{2017ApJ...835...77M}, it is only possible to get a good agreement between the measured and synthetic photometry by choosing a low extinction of E(B-V)=0.8 \citep[similar to extinction measurements near Cas A presented in][]{2015ApJ...810...25G}.
This is in contrast to the extinction of \citet{2017MNRAS.465.3309D}, which might suggest that either the extinction is an overestimate or that the sample consists of foreground stars with less extinction.

\begin{figure}[h!]
 \centering
	 \includegraphics[width=\hsize]{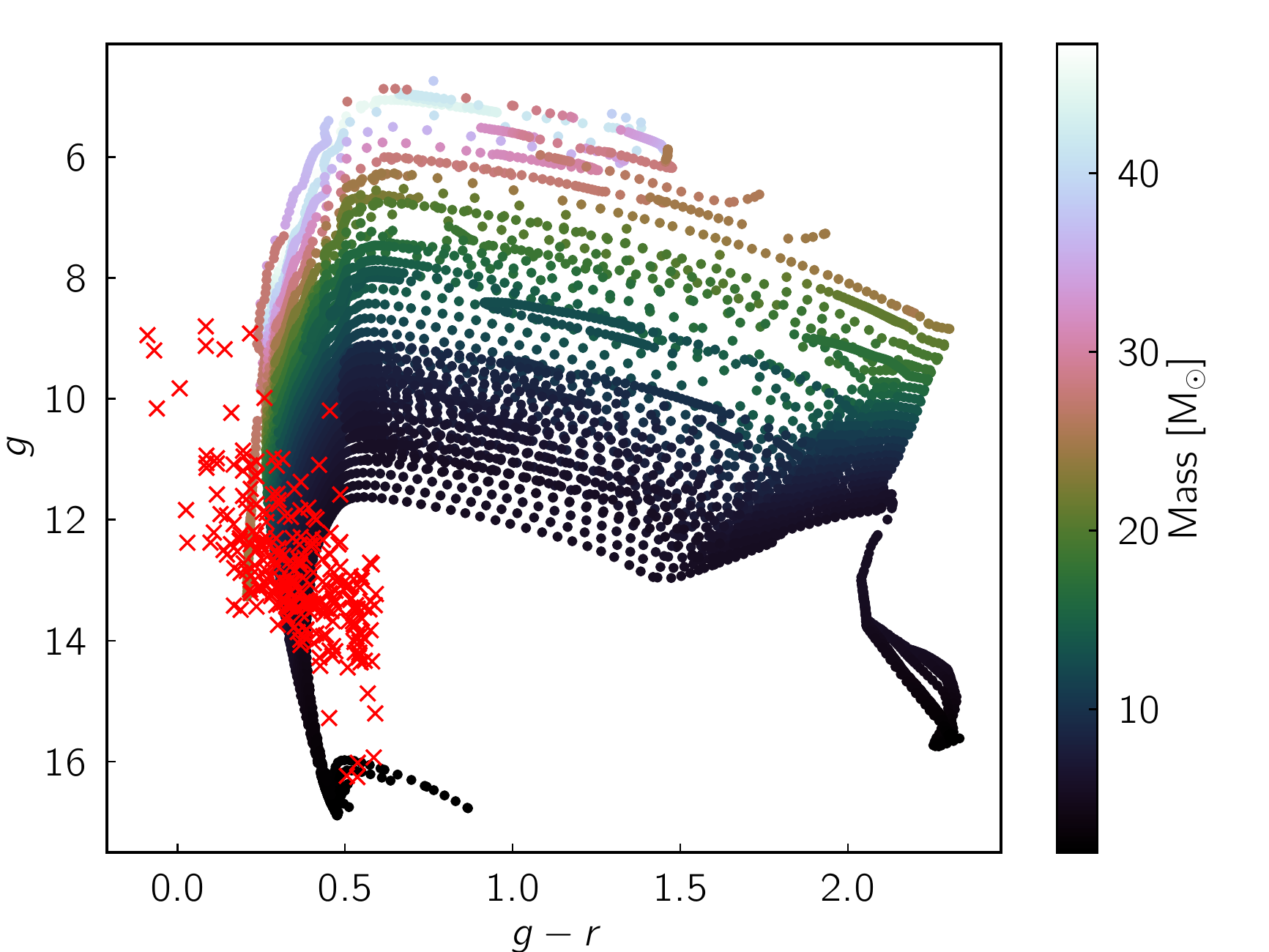}
         \caption{Comparison of the \gls{panstarrs} photometry of $\approx300$ young stars with \textsc{parsec} isochrones \citep{2017ApJ...835...77M} between 6.6 -- 8 $\log_{10}{\textrm{yr}}$.}
       \label{fig:isochrone}
\end{figure}

Figure~\ref{fig:isochrone} demonstrates that these stars are consistent with a distance of 3.4~kpc and shows that there are several candidate stars that (at the distance of 3.4~kpc) might have masses above $30\msun$ which might be able to shed enough of their envelope to explode as a \sniib.

\end{appendix}

\end{document}